\documentclass{article}
\usepackage{arxiv}
\usepackage{graphicx,amsmath,amsfonts,amscd,amsthm,amssymb,pstricks}
\usepackage{color}
\usepackage{dcolumn}
\usepackage{bm}
\usepackage{longtable}
\usepackage{mathrsfs}
\usepackage{textcomp}
\usepackage{esvect}
\usepackage{float}
\usepackage[USenglish,british]{babel}
\usepackage{environ}
\usepackage{xifthen}
\usepackage{xargs}		
\usepackage{hyperref}
\usepackage[separate-uncertainty = true]{siunitx}
\usepackage{physics,mathtools}
\usepackage{multirow}
\usepackage{comment}
\usepackage{enumerate}
\usepackage{appendix}
\usepackage{tikz}	
\usetikzlibrary{arrows, calc, matrix, patterns, decorations.markings, shapes, decorations.pathmorphing, shadows.blur}
\usepackage[utf8]{inputenc}
\usepackage{enumitem}

\newcommand{\av}[1]{\langle #1 \rangle}
\newcommand{\ie}{i.e.\ }

\newcommand{\eg}{e.g.\ }

\newcommand{\ham}{\mathcal{H}}

\newcommand{\eps}{\varepsilon}

\usepackage{xparse}
\newcommand{\phia}{\phi_\mathrm{a}}
\newcommand{\phib}{\phi_\mathrm{b}}
\newcommand{\Ja}{J_\mathrm{a}}
\newcommand{\Jb}{J_\mathrm{b}}

\newcommand{\ii}{I_{\mathrm{i}}}
\newcommand{\Ii}[1]{I_{#1,\mathrm{i}}}
\newcommand{\If}[1]{I_{#1,\mathrm{f}}}

\newcommand{\alfaa}{\alpha_\mathrm{aa}}
\newcommand{\alfab}{\alpha_\mathrm{ab}}
\newcommand{\alfbb}{\alpha_\mathrm{bb}}
\newcommand{\halfaa}{\hat{\alpha}_\mathrm{aa}}
\newcommand{\halfab}{\hat{\alpha}_\mathrm{ab}}
\newcommand{\halfbb}{\hat{\alpha}_\mathrm{bb}}
\def\block(#1,#2)#3{\multicolumn{#2}{c}{\multirow{#1}{*}{$ #3 $}}}
\allowdisplaybreaks[1]

\newcommand{\Ac}{A_\mathsf{ac}}
\newcommand{\Ae}{A_\mathsf{8}}
\newcommand{\AD}{A_\mathsf{D}}
\newcommand{\Ai}{A_\mathrm{i}}
\newcommand{\Af}{A_\mathrm{f}}

\title{Hamiltonian theory of the crossing of the \texorpdfstring{$2 Q_x -2 Q_y=0$}{} nonlinear coupling resonance}

\author{A. Bazzani\\
Dipartimento di Fisica e Astronomia, Universit\`a di Bologna and INFN Bologna, via Irnerio 46, Bologna, Italy\\
\And
F. Capoani\\
Dipartimento di Fisica e Astronomia, Universit\`a di Bologna and INFN Bologna, via Irnerio 46, Bologna, Italy\\
Beams Department, CERN, Esplanade\ des Particules 1, 1211 Geneva 23, Switzerland\\
\And
M. Giovannozzi\thanks{Corresponding author: massimo.giovannozzi@cern.ch}\\
Beams Department, CERN, Esplanade\ des Particules 1, 1211 Geneva 23, Switzerland}

\begin{document}
\maketitle

\begin{abstract}
In a recent paper, the adiabatic theory of Hamiltonian systems was successfully applied to study the crossing of the linear coupling resonance, $Q_x-Q_y=0$. A detailed explanation of the well-known phenomena that occur during the resonance-crossing process, such as emittance exchange and its dependence on the adiabaticity of the process was obtained. In this paper, we consider the crossing of the resonance of nonlinear coupling $2 Q_x -2 Q_y = 0$ using the same theoretical framework. We perform the analysis using a Hamiltonian model in which the nonlinear coupling resonance is excited and the corresponding dynamics is studied in detail, in particular looking at the phase-space topology and its evolution, in view of characterizing the emittance exchange phenomena. The theoretical results are then tested using a symplectic map. Thanks to this approach, scaling laws of general interest for applications are derived.
\end{abstract}


\section{Introduction}
In a recent paper~\cite{PhysRevAccelBeams.24.094002}, the Hamiltonian theory of crossing the resonance $Q_x-Q_y=0$, also called the linear coupling resonance, was presented, and was successfully applied to explain the details of the phenomena that occur during resonance crossing. We recall that the transverse emittances can be exchanged when the resonance is traversed (see Refs.~\cite{Metral:529690,PhysRevSTAB.10.064003} and references therein) and, what is even more important for potential applications, an adiabatic parameter can be defined to qualify the resonance-crossing process. The Hamiltonian theory allows such a parameter to be defined in a natural way~\cite{PhysRevAccelBeams.24.094002}, in contrast to other approaches that have recently been proposed~\cite{PhysRevAccelBeams.23.044003}. It is worth mentioning that the mathematical framework for these studies is the theory of adiabatic invariance for Hamiltonian systems (see \eg Refs.~\cite{NEISHTADT198158,Arnold:937549}). We recall that the key results presented in Ref.~\cite{PhysRevAccelBeams.24.094002} rely mainly on the observation that the Hamiltonian describing the crossing of the resonance remains analytic at the crossing value $Q_x-Q_y=0$, which grants special features to the emittance exchange process, such as the exponential dependence of the exchange rate on the adiabatic parameter. 

The approach used to study the linear coupling resonance has recently been applied to the problem of crossing a generic 2D non-linear resonance~\cite{our_paper7}. This process leads to the possibility of using such a resonance-crossing process to share transverse emittances in a way that depends on the coefficients $m$ and $n$ of the resonance used, that is, $m Q_x-n Q_y = 0$. We recall that the crossing of a 1D nonlinear resonance has previously been studied in view of providing a means to perform Multi-Turn Extraction at the CERN Proton Synchrotron (PS)~\cite{PhysRevLett.88.104801,PhysRevSTAB.7.024001,PhysRevSTAB.10.034001,PhysRevSTAB.12.024003,PhysRevE.89.042915,PhysRevAccelBeams.20.121001}, and which has become the operational means to transfer high-intensity proton beams for the fixed-target physics program at the CERN Super Proton Synchrotron (SPS)~\cite{PhysRevSTAB.9.104001,PhysRevSTAB.12.014001,Borburgh:2137954,PhysRevAccelBeams.20.014001,PhysRevAccelBeams.20.061001,PhysRevAccelBeams.22.104002,Vadai:2702852}. 
In this context, it is interesting to study the crossing of the nonlinear coupling resonance, namely $2Q_x-2Q_y=0$, which is a natural extension to the nonlinear case of the study of the linear coupling resonance. This resonance, also known as the Montague resonance~\cite{Montague:275769} when connected to space-charge effects (see a selection of the literature on this topic in Refs.~\cite{Metral:603510,Hofmann:EPAC04-WEPLT053,Metral:EPAC04-WEPLT029,Hofmann:PAC05-MOPC003,Hofmann:AIP,Hofmann:EPAC06-THPCH006,PhysRevSTAB.9.054202,Qiang:IPAC12-WEPPR011,PhysRevSTAB.16.084201,LeeMontague} and references therein). Our Hamiltonian approach allows the process under study to be generalized, without considering the actual source of the resonance excitation, \ie nonlinear magnetic fields, or space-charge effects, thus retaining only the essential elements characterizing the dynamics of the Hamiltonian system and of the resonance crossing. Our goal is two-fold: to characterize in detail the emittance-exchange process in the non-linear case; to study the impact of a partial resonance crossing, in particular in terms of halo generation, which is an important aspect for any application. 

In Section~\ref{sec:hammod} the Hamiltonian model is introduced and discussed in detail. In particular, the analysis of the phase-space topology is carried out, as well as of the variation of the area of the phase-space regions, which is an essential aspect for the analysis of resonance-crossing phenomena. A map model is introduced in Section~\ref{sec:mapmod} in view of performing the detailed numerical simulations discussed in Section~\ref{sec:results} and used to confirm the processes studied with the Hamiltonian model. Some conclusions are drawn in Section~\ref{sec:conc}. The details of the derivation of the Hamiltonian model are presented in Appendix~\ref{app:derivation}, while Appendix~\ref{app:nf} computes the Normal-Form Hamiltonian.
\section{The Hamiltonian model and its dynamics} \label{sec:hammod}
The starting point of our analysis is the Hamiltonian of a focusing channel in which a pseudo-octupolar term is added to excite the resonance $(2,-2)$. The Hamiltonian depends explicitly on time via $\lambda=\epsilon t$, $\epsilon \ll 1$, with $\epsilon$ being the parameter that describes the resonance-crossing process. The resonance frequencies $\omega_x(\lambda), \omega_y(\lambda)$ are also time dependent. Introducing the parameter $\delta(\lambda)=\omega_x(\lambda)-\omega_y(\lambda)$, to represents the (varying) distance from the resonance, the prototype Hamiltonian can be written as 
\begin{equation}
\begin{split}
    \ham (\phia, & \Ja, \Jb) = \delta (\lambda) \Ja + \frac{1}{2} \alfaa \Ja^2 + \alfab \Ja \Jb + \\
    + & G \Ja (\Jb-\Ja) \cos 2\phia \, .
\end{split}
\label{eq:ham}
\end{equation}
The various steps leading to this Hamiltonian are reported in Appendix~\ref{app:derivation}, and we remark that the terms $\alfaa, \alfab$ are related to the amplitude-detuning terms. Detailed analysis of the properties of this Hamiltonian~\eqref{eq:ham} is fundamental to understand the phenomenon of emittance exchange. With this in mind, it is useful to introduce the rescaled coordinates $J=\Ja/\Jb$, $\phi=\phia$, to give a new Hamiltonian $\mathcal{H}=G \Jb^2\mathcal{H}'$, with
\begin{equation}
    \mathcal{H'}(\phi,J) = \eta J + \alpha J^2 + J (1-J)\cos 2\phi \, ,
    \label{eq:hamf}
\end{equation}
which depends only on two parameters, defined as
\begin{equation} 
\eta=\frac{\delta+\alfab\Jb}{G\Jb},\qquad  \alpha=\frac{\alfaa}{2 \, G} \, ,
\label{eq:transformation}
\end{equation}
where $\eta$ depends on $\lambda$ due to the presence of the term $\delta$.

It is worth pointing out that whenever space-charge effects are considered, the strength of the resonance might become a time-dependent quantity, \ie $G \to G(t)$. In this case, the Hamiltonian~\eqref{eq:ham} describes the crossing of resonance whose strength varies with time according to the physical process under consideration, whereas the resonance-crossing process is controlled independently. On the other hand, by applying the rescaling that brings the Hamiltonian~\eqref{eq:ham} into the form~\eqref{eq:hamf}, the Hamiltonian represents a system in which the resonance strength is constant in time, but the term $J^2$ varies with time, which corresponds to a case in which the amplitude detuning is time dependent. Note, however, that in this case the resonance-crossing process depends on the way the resonance strength varies over time. In this respect, the first Hamiltonian seems more useful as it keeps separate the resonance crossing process and the variation of the resonance strength. The techniques described in the rest of the paper can also be used in the case of a time-dependent $G$ factor, by applying them to the frozen system, \ie with constant $\lambda$ and $t$ values. 

Conditions on the linear actions $I_x, I_y$ in physical coordinates, namely $I_x\ge 0$ and $I_y\ge 0$, are reflected in the inequality $0\le \Ja\le \Jb$, which confines the motion to an \textit{allowed circle}~\cite{PhysRevLett.110.094801,our_paper7}, and which in rescaled coordinates constrains $J$ to the range $[0,1]$. We remark that in the limit $\Jb \to 0$, $\eta \to \infty$, implying that those conditions cross the resonance in a non-adiabatic way. Moreover, it is immediately visible that the following transformation rules 
\begin{equation}
    \eta\to -\eta,\qquad \alpha\to -\alpha,\qquad \phi\to \phi+\frac{\pi}{2}
    \label{eq:symmetry0}
\end{equation}
map $\ham' \to -\ham'$. This symmetry will be used for the classification of possible phase-space topologies. 
\subsection{Fixed points}

The analysis of the fixed points of the Hamiltonian~\eqref{eq:hamf} starts from the equations
\begin{equation}
\begin{dcases}
    \dot \phi &= 2(\alpha-\cos 2\phi)J + \eta + \cos 2\phi=0, \\
    \dot J    &= 2J(1-J)\sin 2\phi=0 \, ,
\end{dcases}
\end{equation}
where the second equation has solutions for $\phi=k\pi/2$, $k\in\mathbb{Z}$, or $J=0$ and $J=1$.

Possible fixed points are given by the following:
\begin{center}
\begin{tabular}{ccc}
$\phi=0,\,\pi$, & $J=\dfrac{1+\eta}{2(1-\alpha)},$ & if $\alpha \neq 1$ \\
& & \\
$\phi=\dfrac{\pi}{2},\,\dfrac{3\pi}{2},$ & $J=\dfrac{1-\eta}{2(1+\alpha)},$ & if $\alpha \neq -1$ \\
& & \\
$J=1$, & $\begin{array}{rl} \phi_{1,\pm} &= \pm\dfrac{1}{2}\acos(2\alpha+\eta)\\ \phi_{2,\pm} &= \pi \pm\dfrac{1}{2}\acos(2\alpha+\eta) \end{array} $ ,  & if $\vert 2\alpha+\eta \vert \leq 1$; \\
& & \\
$J=0$ \, ,
\end{tabular}
\end{center}
whose existence requires that $J \in [0,1]$, which corresponds to
\begin{center}
\begin{tabular}{ccc}
& & \\
$\phi=0,\,\pi$, & $\begin{array}{rcl} 
\eta \ge -1 & \text{and} & \alpha < 1 \\ 
& \text{or} & \\
\eta \le -1 &\text{and} & \alpha > 1 \end{array} $  & and $2\alpha+\eta \le 1$; \\
& & \\
$\phi=\dfrac{\pi}{2},\,\dfrac{3\pi}{2},$ & $\begin{array}{rcl} 
\eta \le 1 & \text{and} & \alpha > -1\\
& \text{or} & \\
\eta \ge 1 & \text{and} & \alpha < -1 \end{array}$  & and $2\alpha+\eta \ge -1$. \\
& & \\
\end{tabular}
\end{center}
We note that the first group of fixed points lies on the $X$ axis, the second group on the $Y$ axis, and the third group on the border of the allowed circle. Furthermore, the first two groups of fixed points coincide with those of the third group whenever $2\alpha + \eta \pm 1 =0$.

The stability type of these fixed points is obtained by considering the determinant of the Hessian matrix of the Hamiltonian, namely
\begin{equation}
    H = \begin{pmatrix}
    2(\alpha-\cos 2\phi) & \phantom{00}& 2(2J-1)\sin 2\phi \\
    & & \\
    2(2J-1)\sin 2\phi    & & 4J(J-1)\cos 2\phi
    \end{pmatrix}\,,
\end{equation}
whose analysis is summarized in Table~\ref{table:Hessian}.
\begin{table}[htb]
    \centering
    \caption{Analysis of the stability type of the fixed points of the Hamiltonian~\eqref{eq:hamf}}
\begin{tabular}{cccc}
\hline 
& & & \\
Solution & \multicolumn{3}{c}{Stability type} \\
          & stable & \phantom{000}& unstable \\
& & & \\
\hline
& & & \\
$\phi=0,\,\pi$ & $\displaystyle\begin{matrix} \alpha \le 1\\ \text{and} \\
-1<\eta<1-2\alpha \end{matrix}$ & &  $\displaystyle\begin{matrix} \alpha \ge 1 \\ 
\text{and} \\
1-2\alpha<\eta<-1 \end{matrix}$ \\ 
& & & \\
$\phi=\dfrac{\pi}{2},\,\dfrac{3\pi}{2}$ & $\displaystyle\begin{matrix} \alpha\ge -1\\ \text{and} \\
-1-2\alpha<\eta<1 \end{matrix}$ & &  $\displaystyle\begin{matrix} \alpha \le -1 \\  
\text{and} \\
1<\eta<-1-2\alpha \end{matrix}$ \\
& & & \\
$J=1$ & never & & always \\
$J=0$ & $\vert \eta \vert >1$ & & $\vert \eta \vert <1$ \\
& & & \\
\hline
\end{tabular}
\label{table:Hessian}
\end{table}

To determine the stability of the fixed point when $J=0$, it is convenient to transform the Hamiltonian into Cartesian coordinates, which are defined as $X=\sqrt{2J}\cos\phi$, $Y=\sqrt{2J}\sin\phi$, allowing the Hamiltonian~\eqref{eq:hamf} to be written in the alternative form:
\begin{equation}
\begin{split}
    \mathcal{H'}(X,Y) &=  \frac{\eta}{2} \left ( X^2+ Y^2 \right ) + \frac{\alpha}{4} \left ( X^2+ Y^2 \right )^2\\ 
                      &\qquad + \frac{1}{4} \qty(2 -  X^2 - Y^2) \qty(X^2- Y^2) \, ,
\end{split}
    \label{eq:hamCart}
\end{equation}
from which it is clear that the transformations $X \to -X$ or $Y \to -Y$ leave the Hamiltonian invariant, and that the dynamics are constrained within the circle $X^2 + Y^2 \leq 2$. For $X=Y=0$, the determinant of the Hessian matrix is $\det H = \eta^2-1$. The fixed point is therefore stable if $\vert \eta \vert >1$ and unstable if $\vert \eta \vert <1$. 

\begin{figure}
    \centering
    \includegraphics[trim=5truemm 5truemm 0truemm 0truemm,width=0.37\columnwidth,clip=]{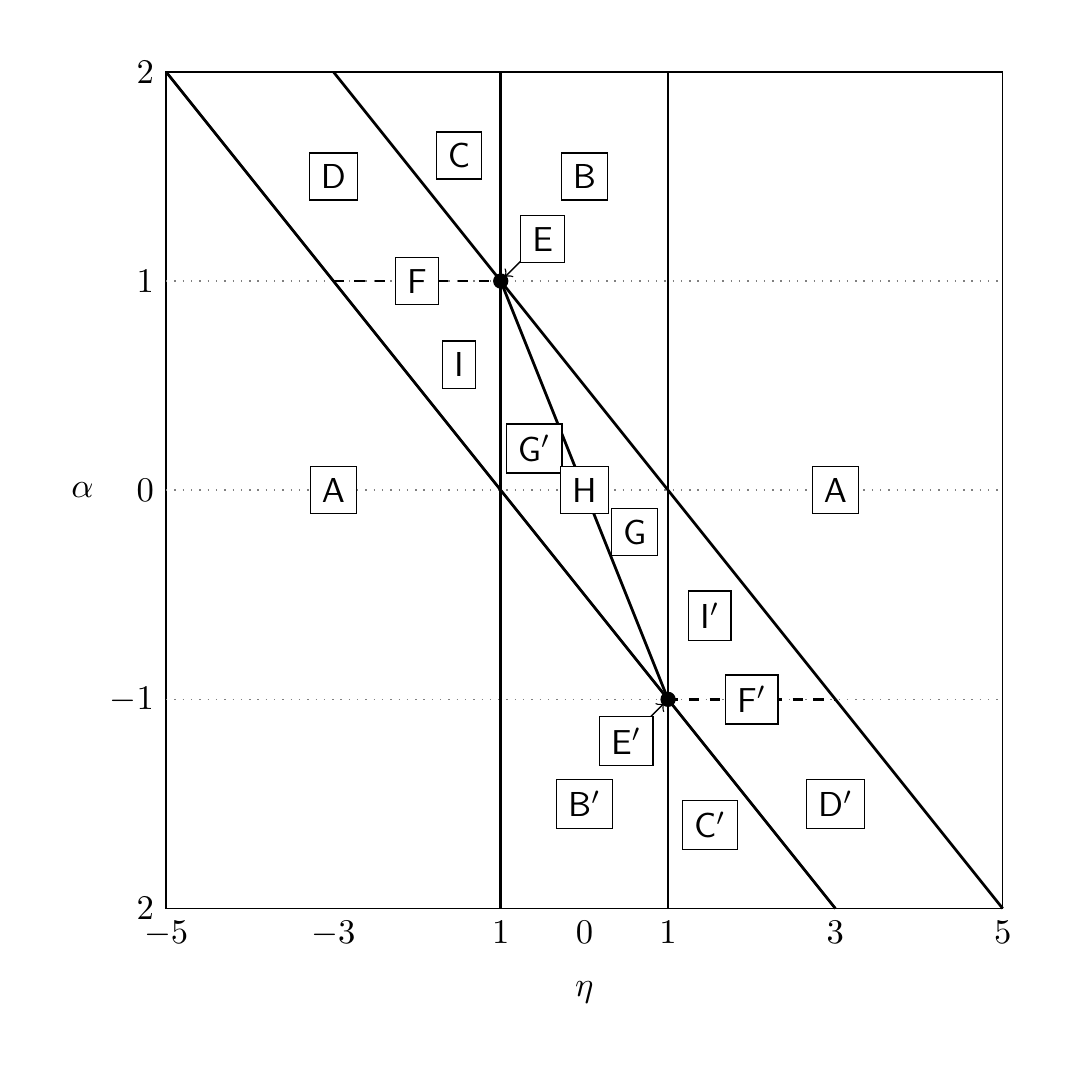}
    \caption{Diagram of the possible phase-space topologies as a function of $\eta$ and $\alpha$. The letters refer to the portraits shown in Fig.~\ref{fig:phsp22}, and the prime indicates a rotation by $\pi/2$ with respect to the configuration shown in Fig.~\ref{fig:phsp22}.}
    \label{fig:eta_alpha_diagram}
\end{figure}

A summary of the possible phase-space topologies is shown in Fig.~\ref{fig:eta_alpha_diagram} in the space $\eta$ and $\alpha$. Note that in case $G=G(t)$ both $\eta$ and $\alpha$ vary with time as $1/G(t)$, describing the straight line $2 \alpha = \alfaa \Jb /(\delta + \alfab \Jb) \eta $.
\subsection{Separatrices} \label{sec:sep}

\begin{figure}
    \centering
    \includegraphics[width=.37\columnwidth]{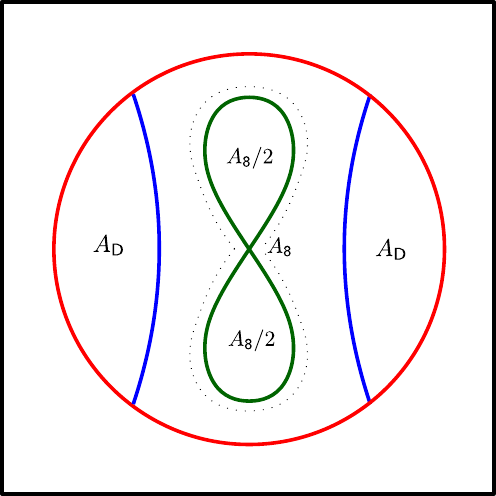}
    \includegraphics[width=.37\columnwidth]{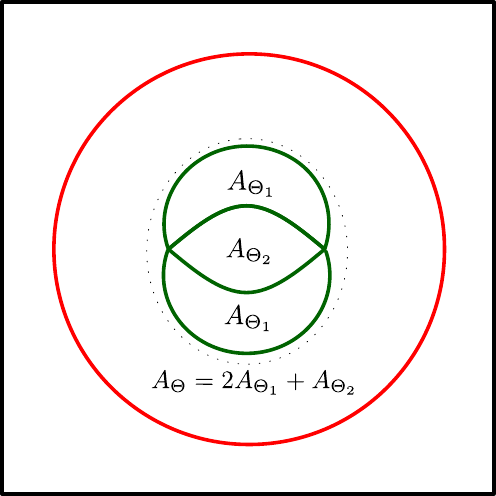}
    \caption{Definition of the names of the various phase-space regions in coordinates $(\sqrt{2J}\cos\phi, \sqrt{2J}\sin\phi)$, corresponding to the topologies, shown in Fig.~\ref{fig:phsp22}, $\mathsf{G}$ (top plot) and $\mathsf{C}$ (bottom plot). Note that the area of the allowed circle, \ie the region inside the red circle, is indicated by $\Ac$. The nomenclature has been introduced in Section~\ref{sec:sep}.}
    \label{fig:areas_details}
\end{figure}
The phase space of Eq.~\eqref{eq:hamf} is divided into different regions by separatrices, whose expressions are computed by solving the equation $\ham(\phi,J)=\ham(\phi_\mathrm{u},J_\mathrm{u})$ for each unstable fixed point (UFP) with co-ordinates  $(\phi_\mathrm{u},J_\mathrm{u})$.

We start from the fixed point $J=0$, which is unstable for $\vert \eta \vert <1$. The equation for the orbit passing through the origin is solved by
\begin{equation}
J(\phi)= \frac{\cos 2\phi + \eta}{\cos 2\phi-\alpha}\,,
\end{equation}
which describes a figure-of-eight in $(\sqrt{2J}\cos\phi, \sqrt{2J}\sin\phi)$ coordinates (see Fig.~\ref{fig:areas_details}, top) passing through $J=0$, and with tangential lines at the origin at angles $\pi/2\pm 1/2\acos \eta$.

The total area $\Ae(\eta)$ of this figure-of-eight is given by the integral
\begin{equation}
\begin{split}
        \Ae & =\int_{\Gamma} \dd\phi\, \frac{\cos 2\phi + \eta}{\cos 2\phi-\alpha} \\
        & = \eta \int_{\Gamma} \dd\phi\, \frac{1}{\cos 2\phi-\alpha} + \int_{\Gamma} \dd\phi\, \frac{\cos 2\phi}{\cos 2\phi-\alpha} \\
        & = \eta I_0 + I_1\,,
\end{split}
\end{equation}
where $\Gamma$ is the domain in $\phi$ for which $J(\phi)>0$ and $I_0, I_1$ are appropriate constants that can be computed numerically. It is indeed possible to compute the integral analytically, but this is not needed in this case as we are only interested in its behavior with respect to $\eta$. When studying the function $1/(\cos 2\phi - \alpha)$, it turns out that 
\begin{equation*}
 I_0<0 \;\; \text{if} \;\; \left\{\begin{array}{rcl}
        \alpha\ge 1, &  & \\
        -1\le \alpha<1 & \text{and} & -\alpha\le \eta \le 1\\
        \end{array}\right .   
\end{equation*}
and
\begin{equation*}
 I_0>0 \;\; \text{if} \;\; \left\{\begin{array}{rcl}
        \alpha\le -1, &  & \\
        -1\le \alpha < 1 & \text{and} & -1 \le\eta\le -\alpha\,.  \\
        \end{array}\right .   
\end{equation*}

When $I_0<0$, $\Ae$ is a non-increasing function of $\eta$, whereas for $I_0>0$, $\Ae$ is a non-decreasing function of $\eta$.

For $\vert \alpha \vert <1$, $\Ae$ is bounded in the interval $[0,\pi/2]$, while for $\vert \alpha \vert >1$ it reaches its maximum at $\eta=-\mathrm{sgn}(\alpha)$ with a value of
\begin{equation}
    \Ae^\text{max}=2\pi\qty(1-\frac{\sqrt{\alpha^2-1}}{\alpha+1})\,.
\end{equation}

Concerning the UFPs, the equation of the separatrix has two solutions, namely  $J=1$, \ie the border of the allowed circle, and 
\begin{equation}
    J(\phi)=\frac{\alpha + \eta}{\cos 2\phi - \alpha}\,.
\end{equation}
This solution is acceptable when $0<J(\phi)<1$, and represents two curves, each one connecting a pair of the four UFPs with $J=1$, which are found for $\phi_\mathrm{u}=1/2\acos(-2\alpha-\eta)$. Each curve is called a \emph{coupling arc}. We define the area of each of the two symmetrical regions delimited by these arcs and by the allowed circle as $\AD(\eta)$ (see Fig.~\ref{fig:areas_details}, top). The value of $\AD$ is given by the integral of $1-J(\phi)$
\begin{equation}
    \begin{split}
    \AD & = \eta \int_{\phi_{1,+}}^{\phi_{2,-}}\frac{\dd \phi}{\alpha - \cos 2\phi} + \int_{\phi_{1,+}}^{\phi_{2,-}}\dd\phi\,\frac{2\alpha - \cos 2\phi}{\alpha-\cos 2\phi} \\
    & = \eta I_2 + I_3 \,,
    \end{split}
\end{equation}
where $I_2, I_3$ are appropriate constants that can be computed numerically.

As for the case of $\Ae$, we are interested to study how $\AD$ varies as a function of $\eta$, which depends on the sign of $I_2$. We have the following 
\begin{equation*}
 I_2>0 \;\; \text{if} \;\; \left\{\begin{array}{rcl}
        \alpha\ge 1, &  & \\
        -1\le \alpha<1 & \text{and} & -1-2\alpha\le \eta \le -\alpha\\
        \end{array}\right .   
\end{equation*}
and
\begin{equation*}
 I_2<0 \;\; \text{if} \;\; \left\{\begin{array}{rcl}
        \alpha\le -1, &  & \\
        -1\le \alpha < 1 & \text{and} & -\alpha \le\eta\le 1-2\alpha\,.  \\
        \end{array}\right .   \, .
\end{equation*}
For the case of $\Ae$, $\AD$ is monotonically non-increasing if $I_2<0$ and monotonically non-decreasing if $I_2>0$.

Finally, UFPs might exist either for $\phi=0,\pi$ or for $\phi=\pi/2,3\pi/2$. We analyze the case $\alpha>1$ and $1-2\alpha<\eta<-1$, where two UFPs are found for $\phi=0$ and $\phi=\pi$ (the case of $\phi=\pi/2,3\pi/2$ is analogous). 

To express the solution, it is useful to define the parameters $a=\alpha-1$ and $u=-(\eta+1)/(2a)$, so that $u \in [0,1]$. Therefore we have 
\begin{equation}
    J(\phi)= \frac{au + \sin^2\phi \pm \sqrt{\sin^2\phi + 2au(1-u)}}{a+2\sin^2\phi}\,.
\end{equation}

The solutions delimit three regions of phase space: an upper and a lower region, both with area $A_{\Theta_1}$, and a central region, with area $A_{\Theta_2}$. The two outer regions each enclose one stable fixed point where $\phi=\pi/2$ or $\phi=3\pi/2$, while the inner region encloses the stable origin. The corresponding phase-space topology with the definition of the areas is shown in the bottom plot of Fig.~\ref{fig:areas_details}.

The area of these regions can be evaluated by the integrals
\begin{equation}
\begin{split}
    A_{\Theta_1}(u) &= 2\int_0^\pi \dd\phi\,\frac{\sqrt{\sin^2\phi + 2au(1-u)}}{a+2\sin^2\phi} \\
    & \\
    A_{\Theta_2}(u) &= 2\int_0^\pi \dd\phi\, \frac{au + \sin^2\phi - \sqrt{\sin^2\phi + 2au(1-u)}}{a+2\sin^2\phi}\, ,
\end{split}
\end{equation}
where $A_1(u)$ is symmetric with respect to $u=1/2$. We define the area of the entire three-region structure as $A_\Theta = 2A_{\Theta_1}+A_{\Theta_2}$, noting that
\begin{equation}
    A_\Theta(1/2\pm v)+ A_{\Theta_2}(1/2\mp v) = 2\pi
    \label{eq:areas_prop}
\end{equation}
for any $v\in[0,1/2]$.

It is possible to show that $A_{\Theta_2}$ grows as $u$ is increased (and $\eta$ decreased). On the other hand, $A_{\Theta_1}$ reaches its maximum at $u=1/2$. $A_\Theta$ is also monotonic. At $u=0$, $A_\Theta=\Ae^\text{max}$, while for $u=1$, $A_\Theta=2\pi$, and the three-region structure covers the entire allowed circle, whose area is given by $\Ac=2\pi$.
\subsection{Classification of the phase-space topology}
\begin{figure*}
\includegraphics[trim=5truemm 0truemm 5truemm 0truemm,width=\textwidth,clip=]{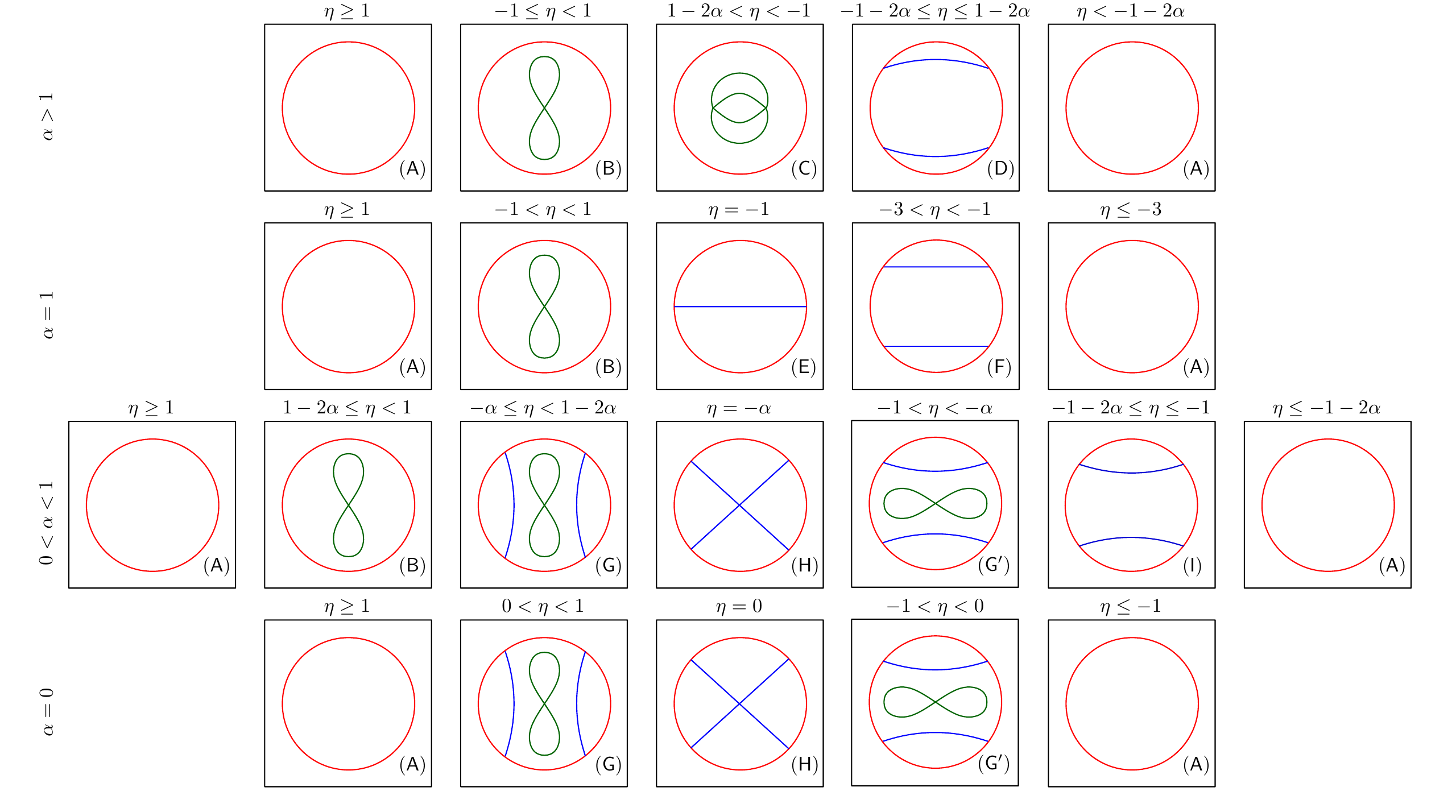}
\caption{Possible phase-space topologies for the Hamiltonian~\eqref{eq:hamf} as a function of $\alpha$ (vertical direction) and $\eta$ (horizontal direction), using Cartesian co-ordinates $(\sqrt{2J}\cos\phi, \sqrt{2J}\sin\phi)$. Note that $\eta$ is decreasing from left to right, as we describe a process where $\eta$ starts at $\eta\gg 1$ and ends at $\eta\ll 1$. The border of the allowed circle is depicted in red, the coupling arcs are in blue, and the figure-of-eight separatrix in green. The letters indicate the topology type marked in Fig.~\ref{fig:eta_alpha_diagram}.}
\label{fig:phsp22}
\end{figure*}

Using information on the existence and type of stability of the fixed points, it is possible to reconstruct all possible phase-space topologies, as shown in Fig.~\ref{fig:phsp22}, as a function of $\eta$ and $\alpha$. Phase-space portraits, represented using Cartesian coordinates, are shown with $\alpha$ along the vertical direction (increasing from bottom to top), and $\eta$ in the horizontal direction (increasing from left to right). In this way, a resonance crossing process occurs according to a sequence of phase-space structures corresponding to a certain row shown in Fig.~\ref{fig:phsp22}. In all cases shown, the initial phase-space structure is the simplest and transforms through different sequences of topologies to a final configuration that is again the simplest possible.

In general, when $\vert \eta \vert \gg 1$ only the fixed point in $J=0$ exists. The origin then becomes unstable and a figure-of-eight appears in which two stable fixed points are present either along the horizontal or the vertical axis. The existence of four fixed points on the border of the allowed circle creates two coupling arcs, each enclosing one stable fixed point.

\begin{figure*}
\includegraphics[width=\textwidth]{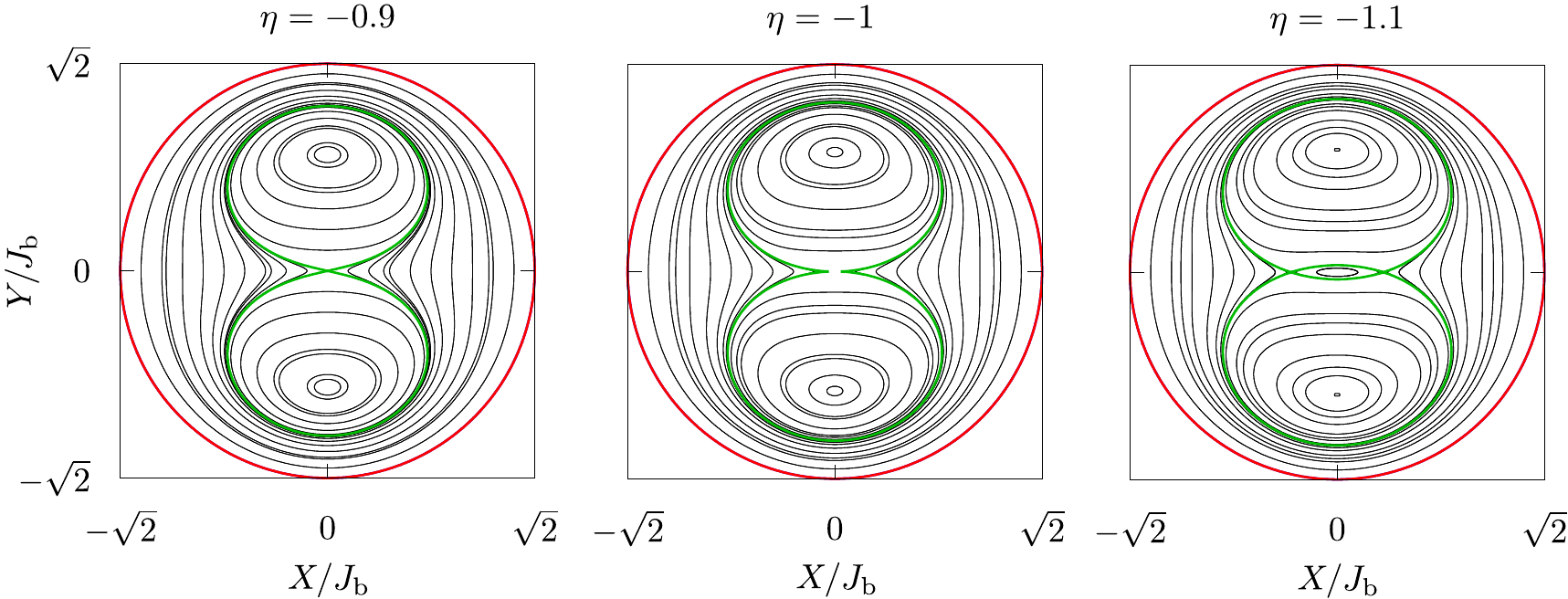}
\caption{Detail of the transition between the phase-space topology $\mathsf{(B)}$ and $\mathsf{(C)}$ shown in Fig.~\ref{fig:phsp22} for $\alpha=2$. We remark that a bifurcation phenomenon occurs at the origin at $\eta=-1$.}
\label{fig:phsp_transition}
\end{figure*}

The case $\alpha=0$ is the simplest with a highly symmetrical phase-space structure, which is mirrored around $\eta=0$. When $0 < \alpha < 1$ a rich combination of phase-space structures appears, which is reduced when $\alpha=1$, although horizontal coupling arcs are generated for the first time. Finally, for $\alpha > 1$ a new structure appears, corresponding to the birth of two unstable fixed points, which were absent for smaller values of $\alpha$. Details of the transition between the phase-space topology $\mathsf{(B)}$ and $\mathsf{(C)}$ in Fig.~\ref{fig:phsp22} are shown in Fig.~\ref{fig:phsp_transition}. The orbits of the system are displayed for $\alpha=2$ and three values of $\eta$ to visualize the transition from the figure-of-eight shaped region, with two stable and one unstable fixed points, to the new structure, which features three stable and two unstable fixed points.  

We underline that in Fig.~\ref{fig:phsp22}, only the phase-space portraits corresponding to $\alpha>0$ are presented, as due to the symmetries of the Hamiltonian under consideration, the case $\alpha<0$ is the same as $\alpha>0$, provided that the sign of $\eta$ is reversed and $\phi$ is shifted by $\pi/2$.

\subsection{Resonance-crossing process}

We analyze the resonance crossing process for the Hamiltonian~\eqref{eq:hamf} when $\delta$ is slowly varied from $\delta \gg 1$ to $\delta \ll -1$, with $\delta \approx \epsilon t$, where $0 < \epsilon \ll 1$ is the parameter that controls the adiabaticity of resonance crossing. Given the link between $\delta$ and $\eta$, the latter is also changed with time. By applying the approach used in Refs.~\cite{PhysRevAccelBeams.24.094002,our_paper7}, it is possible to show that the correct adiabatic parameter is actually given by $\epsilon/(G \Jb^2)^2$, which means that whenever $\epsilon$ is varied, $G$ should be readjusted according to the scaling law $G \approx \sqrt{\epsilon}$ to keep the crossing process invariant. 

We assume that the parameter variation is performed under adiabatic conditions, so that the adiabatic theory for Hamiltonian systems holds~\cite{neish1975,NEISHTADT198158,NEISHTADT1986,an9,an10,NEISHTADT2006158,Neishtadt_2019}. We recall that adiabatic theory describes the evolution of the orbits using the action-angle variables of the frozen Hamiltonian in each region defined by the separatrix curves.  The area enclosed by an orbit, \ie $2\pi$ times the initial action $\ii$, remains almost constant during the variation, until its value coincides with the area of the regions delimited by a separatrix. It then crosses the separatrix and enters into a growing region of phase space with a probability proportional to the time derivative of the area of each region. At the separatrix crossing, the adiabatic theory has to be improved~\cite{NEISHTADT1986}, since the action function is singular and the action value may have small stochastic changes. After the separatrix crossing, the new action value of the orbit corresponds to the area enclosed by the separatrix at the crossing time divided by $2\pi$, and is preserved in the adiabatic approximation.

\begin{figure*}
    \centering
    \includegraphics[width=\textwidth]{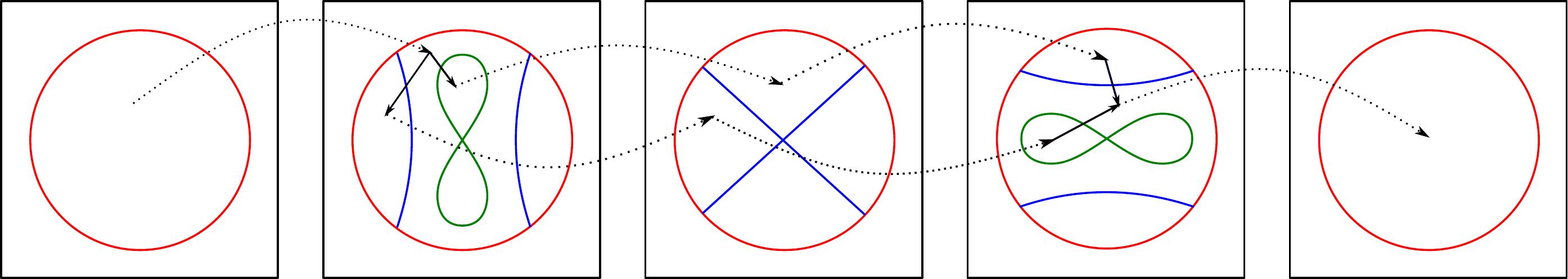}
    \caption{Sketch of the resonance-crossing process ($\eta$ decreases from left to right) for the simple case $\alpha=0$. The dotted arrows indicate the transition between regions that do not imply separatrix crossing, whereas continuous arrows indicate the transition between regions that require separatrix crossing.}
    \label{fig:alpha0_process}
\end{figure*}

Let us follow the evolution of an initial condition during such a resonance crossing process, starting from the simplest case, \ie $\alpha=0$. The sketch of possible phenomena that occur during the resonance-crossing process is shown in Fig.~\ref{fig:alpha0_process}, where the phase-space portraits of Fig.~\ref{fig:phsp22} are depicted and information about possible transitions between the different phase-space regions is included. The dotted arrows indicate the possible transitions that do not imply any separatrix crossing, whereas the continuous arrows indicate transitions between regions that require separatrix crossing. 

At first, when $\eta \ge 1$, the particle has an action $\ii=\Ai/2 \pi$ where $\Ai$ is the area enclosed by the orbit. We remark that $\Ae(\eta)$ and $\AD(\eta)$ reach their maximum values at $\eta=0$, when they degenerate~\footnote{In this case, $\Ae(0)$ is made of the upper and lower circular sectors, and $\AD(0)$ of the left and right circular sectors.}, and that $0\le \Ae \le \pi$, while $0\le \AD \le \pi/2$.

When $\eta$ becomes smaller than $1$, several phase-space regions are generated by the appearance of stable and unstable fixed points and the orbit will be confined in the area outside of $\Ae$ and $\AD$ as long as $\Ae(\eta)\le \Ai \le \Ac - 2\AD(\eta)$. We define $\eta^\ast$ as the value of $\eta$ when one of the two limiting conditions is first met, \ie either $\Ai+2\AD(\eta^\ast)=\Ac$ or $\Ai=\Ae(\eta^\ast)$, where $\Ac$ has been introduced in Fig.~\ref{fig:areas_details}.

In the first case, the particle is trapped in the left lobe, with a new enclosed area $A_1=\AD(\eta^\ast)=(\Ac-\Ai)/2$. When $\eta<0$, the lobe becomes the left half of the new, horizontally-oriented figure-of-eight, which starts shrinking. The particle remains within the lobe until $\eta=\eta^{\ast\ast}$, when $\Ae(\eta^{\ast\ast})/2=A_1$. At that moment, the particle is released out of the separatrix, and is enclosed in the area $\Af=\Ae(\eta^{\ast\ast})=2A_1=\Ac-\Ai$. This value is conserved in the final state, when $\eta\le -1$.

On the other hand, if the initial condition satisfies $\Ai=\Ae(\eta^\ast)$, the particle is initially trapped in the upper lobe of the vertically oriented figure-of-eight, with an enclosed area $A_1=\Ae(\eta^\ast)/2=\Ai/2$. When $\eta<0$, the particle stays in the upper region delimited by the coupling arc. This lobe shrinks, and for $\eta=\eta^{\ast\ast}$, when $\AD(\eta^{\ast\ast})=A_1$, the particle is released from the lobe, with a new orbit area $\Af=\Ac - 2\AD(\eta^{\ast\ast})=\Ac-2A_1=\Ac-\Ai$. In both cases, we have $\Af=\Ac-\Ai$, which means that if the particle had an initial action value $\ii=\Ai/(2\pi)$, it will end up having an action $I_\mathrm{f} = \Af/(2\pi) = (\Ac-\Ai)/2\pi = 1-\ii$ or, transforming back to $I_x$ and $I_y$ coordinates,
\begin{equation}
    \If{x}=\Ii{y},\qquad \If{y}=\Ii{x} \, ,
\end{equation}
corresponding to an emittance exchange mediated by the resonance-crossing process.

The same reasoning applied to the cases $0<\alpha<1$ and $\alpha=1$ yield the same result. We observe that when $1>\eta>-\alpha$, the particle is trapped in the figure-of-eight or in one of the lobes, and is detrapped when $-\alpha>\eta>-1-2\alpha$. 

The analysis of the case $\alpha>1$ is more involved, as three possibilities exist. Let us consider the changes of the phase-space topology in a qualitative way. As $\eta$ decreases, the figure-of-eight appears when $\eta=1$ and grows until it reaches its maximum area at $\eta=-1$, when a bifurcation phenomenon occurs. 
%
%
When $\eta<-1$, a new structure appears (see 
Fig.~\ref{fig:phsp_transition}), which is divided into three parts and with total area $A_\Theta$ that satisfies $\Ae^\text{max}\le A_\Theta \le \Ac$ when $\eta\in[-1,1-2\alpha]$. 

Let us consider the three possible cases. If $\Ai<\Ae^\text{max}$, the particle is trapped in the figure-of-eight when $-1\le \eta\le 1$, enclosing an area $A_1 =\Ai/2$. It then moves into the zone $A_{\Theta_1}$, and as $A_{\Theta_1}>\Ae^\text{max}$ for $\eta\in[1-2\alpha,-1]$, this occurs without crossing any separatrix during that interval. However, for $\eta < 1-2\alpha$, the particle is released back to the center for $\eta=\eta^{\ast\ast}$, when $\AD=A_1=\Ai/2$. Hence, the final area is $\Af=\Ac-2A_1=\Ac-\Ai$, and the emittance exchange holds as seen in the previous cases.

If $\Ai>\Ae^\text{max}$, the first trapping occurs for $\eta\in[1-2\alpha,-1]$. $A_{\Theta_1}$ varies in a symmetrical way with respect to $\eta=1-\alpha$, and its derivative is positive only if $\eta>1-\alpha$. Therefore, defining $\eta^\ast$ as the solution of equation $A_{\Theta_1}(\eta^\ast)=\Ai$, if $\eta^\ast>1-\alpha$ the particle can be trapped according to a probability law~\cite{neish1975}, either in $A_{\Theta_1}$ (with probability $\mathcal{P}_1$) or in $A_{\Theta_2}$ (with probability $1-\mathcal{P}_1$). If $\eta^\ast<1-\alpha$, the particle is trapped in $A_{\Theta_2}$.

A particle trapped at $\eta=\eta^\ast$ in $A_{\Theta_1}$ has the orbit area $A_{\Theta_1}(\eta^\ast)$. As $A_{\Theta_1}$ grows and then shrinks back to the same area when $\eta^{\ast\ast}=1-\alpha-\eta^\ast$, the particle reaches the final area $A_{\Theta_2}(\eta^{\ast\ast})$ and the following relation holds
\begin{equation}
    \begin{split} 
    \Af &= A_{\Theta_2}(\eta^{\ast\ast})\\
        &=A_\Theta(\eta^{\ast\ast})-2A_{\Theta_1}(\eta^\ast) \\
        &= A_\Theta(\eta^\ast+1-\alpha)+A_{\Theta_2}(\eta^\ast)-\Ai\\
        &= 2\pi - \Ai\,.
    \end{split}
\label{eq:exch}
\end{equation}
The last equality is derived from Eq.~\eqref{eq:areas_prop}, and once more an emittance exchange occurs.

The third possibility is that a particle with $\Ai=A_\Theta(\eta)$ at $\eta=\eta^\ast\in[1-2\alpha,-1]$ is immediately trapped in $A_{\Theta_2}$. When $\eta^\ast>1-\alpha$, this occurs with a probability $\mathcal{P}_1$, with probability $1-\mathcal{P}_1$ this does not occur and the previous case applies. On the other hand, when $\eta^\ast<1-\alpha$, trapping in $A_{\Theta_2}$ is guaranteed and the final area is $\Af= A_{\Theta_2}(\eta^\ast)$, or $A_\text{f}=\Ai-2A_{\Theta_1}(\eta^\ast)$, which does not result in emittance exchange. This brings us to an interesting observation. Unlike the case of the $(1,-1)$ linear coupling resonance~\cite{PhysRevAccelBeams.24.094002}, in which the amplitude detuning affects only the adiabaticity of the crossing process, in the case of the $(2,-2)$ nonlinear coupling resonance, the amplitude detuning might perturb the emittance exchange proper and could even prevent it from occurring. Such effect is due to the phase-space topology that, when $\alpha>1$, features structures that may prevent emittance exchange.  

\begin{figure}
    \centering
    \includegraphics[width=0.47\columnwidth]{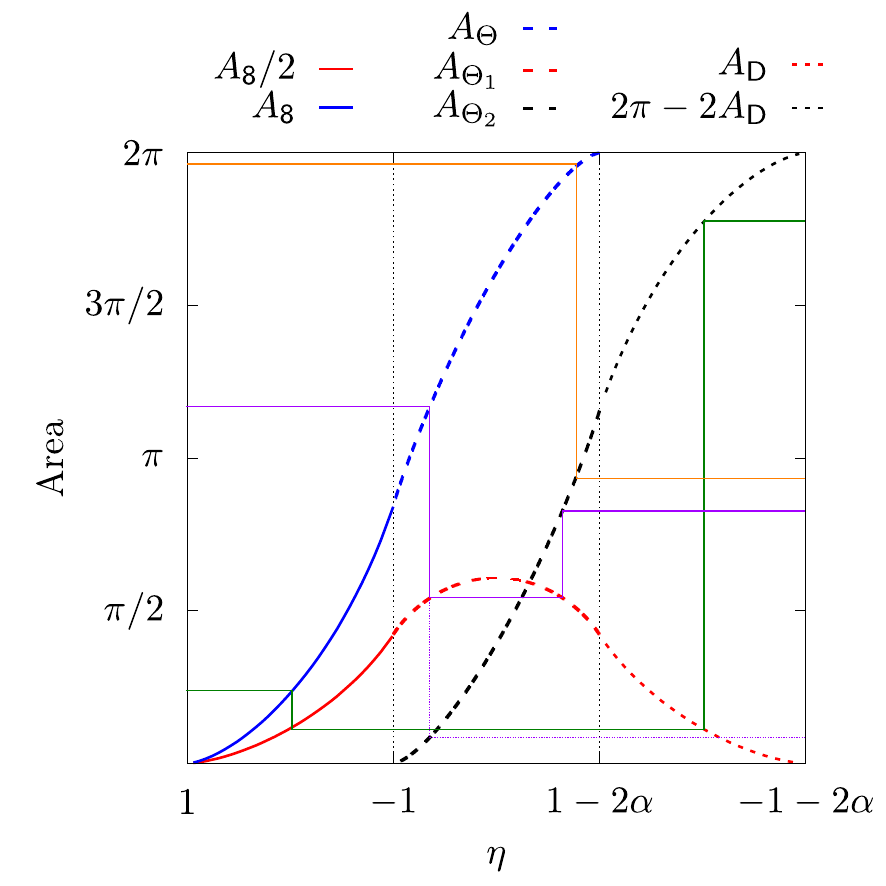}
    \caption{Scheme of the trapping process in the case $\alpha>1$. Thick lines represent the areas of the different phase-space regions as a function of $\eta$ (note the reversed horizontal scale, as we describe a process in which $\eta$ decreases with time). The same color represents a continuity of a region throughout the intervals $1>\eta>-1$, $-1>\eta>1-2\alpha$, $1-2\alpha>\eta>-1-2\alpha$. Thin lines show the evolution of the area with three possible initial conditions. The green line represents a particle that is first trapped in $\Ae$ and then released from $\AD$. The purple line shows (with a bifurcation) the two possible outcomes of a particle trapped either in $A_{\Theta_1}$ and then released, or in $A_{\Theta_2}$ at $\eta^*>-1-\alpha$. Finally, the orange line represents a particle trapped in $A_{\Theta_1}$ when $\eta^*<-1-\alpha$.}
    \label{fig:areas_alpha1}
\end{figure}

The possible processes that occur for $\alpha>1$ are summarized in Fig.~\ref{fig:areas_alpha1}, where the areas of the phase-space regions, computed by means of numerical evaluation of the corresponding integrals, are plotted as functions of $\eta$. The blue curve represents the outer area, $\Ae$ or $A_\Theta$, the red curve represents the area of the region around the stable point in $\phi=\pi/2$, and the black curve represents the area of the region that includes the origin when it is stable. The green line shows the area evolution of a particle that is trapped inside the figure-of-eight region, while the solid purple line shows a particle that is first trapped in $A_{\Theta_1}$. The dotted purple line and the orange line are two particles that are immediately trapped in $A_{\Theta_2}$. Note that the purple line bifurcates since both outcomes are possible for a particle with $\Ae^\text{max}\le \Ai \le A_\Theta(1-\alpha)$.

In summary, when $|\alpha|\le 1$, we expect that due to a resonance-crossing process the initial normalized action $J_\text{i}$ of each particle becomes $1-J_\text{i}$, corresponding to an exchange of the values of $I_x$ and $I_y$. On average, this results in the exchange of emittances $\eps_x=\av{I_x}$ and $\eps_y=\av{I_y}$ after the crossing.

On the contrary, when $|\alpha|>1$, only some initial conditions undergo the action jump $J_\text{i}\to 1-J_\text{i}$ and contribute to emittance exchange. The final efficiency in the exchange of emittances  will therefore depend on the fraction of particles with the right initial conditions. This corresponds to particles with sufficiently small or large $J_\text{i}$, depending on the sign of $\alpha$ and the direction of resonance crossing. The emittance exchange performance in conditions of high-amplitude-detuning is therefore sensitive to the initial distribution, in particular to the initial emittance ratio. 

\section{The map model} \label{sec:mapmod}
The dynamics generated by the Hamiltonian~\eqref{eq:hamq} corresponds to the quasiresonant Normal Form expansion of a 4D Hénon-like map~\cite{Bazzani:262179} with cubic nonlinearity (see Appendix~\ref{app:nf}). This polynomial map simulates the one-turn map of a FODO cell in which a normal octupole, represented as a single-kick element, is located. In Courant-Snyder coordinates, the Hénon-like map reads 
\begin{equation}
\renewcommand\arraystretch{1.6}
\begin{pmatrix}
x'\\p'_x\\ y' \\p'_y
\end{pmatrix} = R(\omega_x,\omega_y) \begin{pmatrix} x &\\ p_x &+ \dfrac{K_3}{6}\beta^2_x\qty(x^3-3\beta xy^2)  \\ y &\\ p_y &+ \dfrac{K_3}{6}\beta^2_x\qty(\beta^2 y^3 -3 \beta x^2 y) \end{pmatrix}\, ,
\label{eq:henon}
\end{equation}
where $R$ is a $4\times 4$, block diagonal matrix, whose blocks are $2\times 2$ rotation matrices of frequencies $\omega_x, \omega_y$. The parameter $\beta=\beta_y/\beta_x$ represents the ratio of the $\beta$-functions at the location of the octupole, and $K_3$ is the normalized octupolar strength, defined as
\begin{equation}
    K_3 = \frac{\ell}{B \rho} \frac{\partial^3 B_y}{\partial x^3} \, 
\end{equation}
where $B\rho$ is the magnetic rigidity, $\ell$ is the length of the octupole, and $B_y$ is the vertical component of the magnetic field.

By computing  the Normal Form expansion of Eq.~\eqref{eq:henon}, we can establish the correspondence between the parameters of the Hamiltonian~\eqref{eq:ham12} and those of the map in the neighborhood of the origin (see Appendix~\ref{app:nf}), namely
\begin{equation}
\begin{split}
    G      & =  \phantom{-}K_3 \frac{\beta_x\beta_y}{4} \, , \\
    \alfaa & = -K_3\qty(\frac{1}{4}\beta_x^2 + \beta_x\beta_y + \frac{1}{4}\beta_y^2)\, ,\\ 
    \alfab & =  \phantom{-}\frac{K_3}{2} \beta_y \qty(\beta_x + \frac{\beta_y}{2})\, , \\
    \alfbb & = -K_3 \frac{\beta_y^2}{4} \,.
\end{split}
\end{equation}

These formulae show that a normal octupole always generates an amplitude-dependent detuning, which is reflected in the values of $\alfaa$ and $\alfab$. It is worth recalling that, thanks to superperiodicity, it is possible to assume that additional octupoles can be installed in the ring lattice so that they only contribute to amplitude-detuning and not to the resonant term. This configuration has been included in our model by changing the rotation frequency of the matrix $R$ in Eq.~\eqref{eq:henon} by introducing action-dependent terms, \ie.
\begin{equation}
\begin{split}
     \omega_x &\to \omega_x + \overline{\alpha_{xx}} I_x + \overline{\alpha_{xy}} I_y \\
     \omega_y &\to \omega_y + \overline{\alpha_{xy}} I_x + \overline{\alpha_{yy}} I_y \,.
     \label{eq:detuning}
\end{split}
\end{equation}

Although the case $\beta=1$ represents a simplification of the general case, it retains some interest. By fixing $\overline{\alpha_{xx}}=\overline{\alpha_{yy}}=0$, the parameter $\overline{\alpha_{xy}}$ can be used to control the value of $\alpha$ in Eq.~\eqref{eq:hamf}. In this case, the computation of the Normal Form on the map of Eq.~\eqref{eq:henon} with the amplitude-dependent rotation frequencies of Eq.~\eqref{eq:detuning} gives
\begin{equation}
    \alfaa=-\frac{3}{2}K_3 - 4\overline{\alpha_{xy}},\qquad \alfab=\frac{3}{4}K_3 + 2\overline{\alpha_{xy}},
    \label{eq:detuning_map}
\end{equation}
and the parameter $\alpha$ of Eq.~\eqref{eq:hamf} is given by
\begin{equation}
    \alpha=-3 - 8\frac{\overline{\alpha_{xy}}}{K_3}\,.
    \label{eq:alpha_map}
\end{equation}

In the numerical simulations using the map of Eq.~\eqref{eq:henon}, and adding the amplitude-dependent terms of Eq.~\eqref{eq:detuning}, we remark that the amplitude-dependent rotation of Eq.~\eqref{eq:detuning} uses the invariants of the frozen system to evaluate $I_z, z=x \, \text{or} \, y$. In a nonlinear system, the linear actions $I_z = 1/2(z^2 + p_z^2)$ are not invariant for the dynamics, and an approximation of the actual invariant can be obtained by using Birkhoff Normal Forms~\cite{Bazzani:262179} (see also Appendix~\ref{app:nf}). For the case of numerical simulations, we computed the correction up to the fourth order using the software described in~\cite{Bazzani:1995vj}. Neglecting this effect would move each particle to a different value of the action at each application of the amplitude-dependent rotation, thus causing a loss of symplecticity of the system.
\section{Results of numerical simulations} \label{sec:results}
%
%
We measure the performance of the emittance exchange process for resonance $(2,-2)$ by computing the evolution of a Gaussian distribution of initial conditions $\rho(I_x,I_y)$ under the dynamics generated by the map of Eq.~\eqref{eq:henon} iterated for $N$ turns, with or without amplitude-detuning terms. We recall that no complete and rigorous results for the adiabatic theory of time-dependent symplectic maps are available. On the other hand, the use of the interpolating Hamiltonian allows one to apply a perturbation approach. In the numerical simulations, $\omega_x$ is kept constant while $\omega_y$ is linearly varied between the initial value {$\omega_{y,\text{i}} = \omega_x + \delta_\text{max}$} and the final value $\omega_{y,\text{f}} = \omega_x - \delta_\text{max}$, to cross the resonance. The initial and final emittances were then compared with the figure of merit $P_\text{na}$, introduced in Ref.~\cite{PhysRevAccelBeams.23.044003}, used to assess how well the emittance exchange occurred. $P_\text{na}$ is defined as
\begin{equation}
    P_\text{na} = 1 -  \frac{ \av{\If{x}} - \av{\Ii{x}}}{\av{\Ii{y}} - \av{\Ii{x}}}
    \label{eq:pna}
\end{equation}
that satisfies $P_\text{na}=1$ when no emittance exchanged occurs and $P_\text{na}=0$ when the exchange is perfect.

The presence of a halo in the final distribution is expected for a small set if initial conditions due to the changes in the adiabatic invariant at separatrix crossing when the time-variation is not perfectly adiabatic~\cite{NEISHTADT198158}. This phenomenon needs to be determined quantitatively by means of numerical simulations to provide a complete assessment of the resonance-crossing process. To this aim, it is possible to define the so-called halo parameter~\cite{PhysRevSTAB.5.124202}:
\begin{equation}
h_z = \frac{\av{z^4}}{\av{z^2}^2}-2 \qquad z=x \; \text{or} \; y\, .
\label{halodef}
\end{equation}

\begin{figure*}[htb]
    \centering
    \includegraphics[width=0.47\columnwidth]{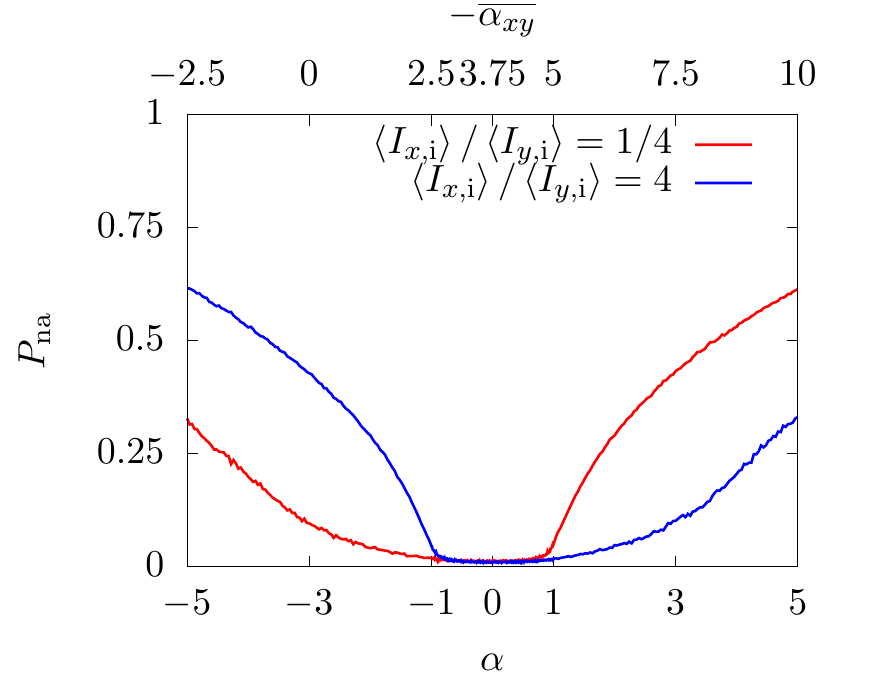}
    \includegraphics[width=0.47\columnwidth]{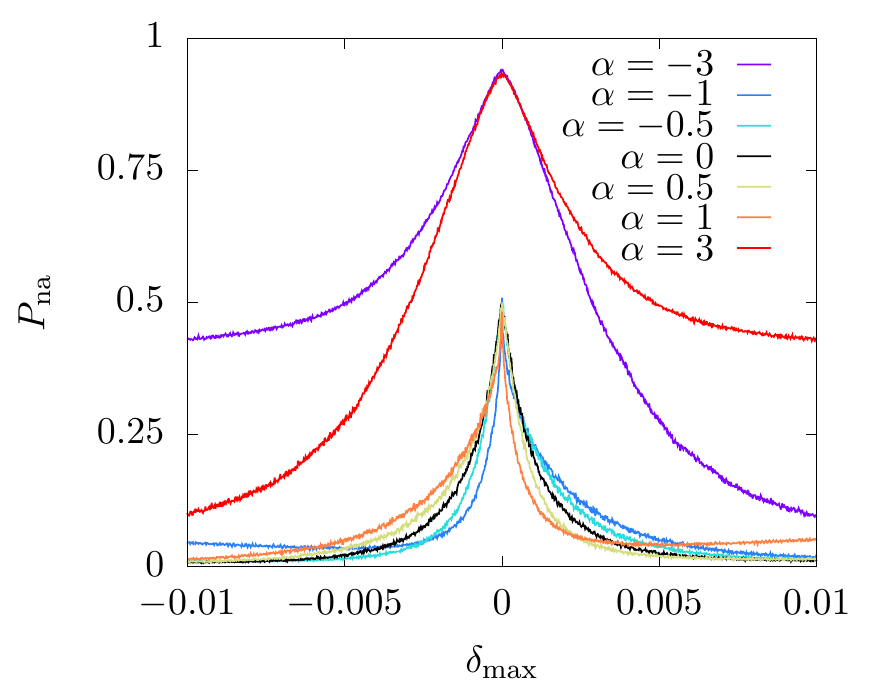}
    \includegraphics[width=0.47\columnwidth]{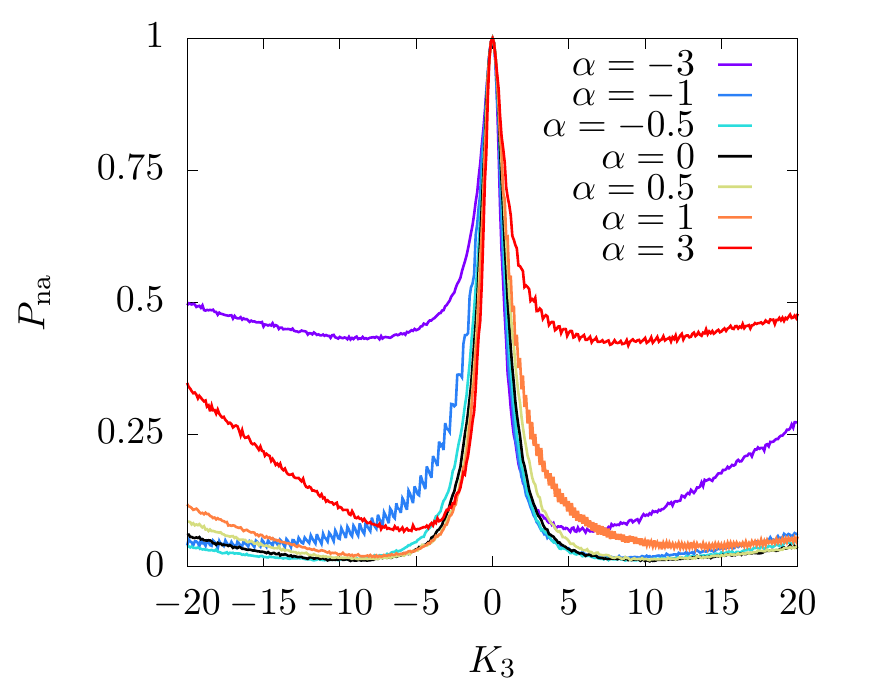}
    \includegraphics[width=0.47\columnwidth]{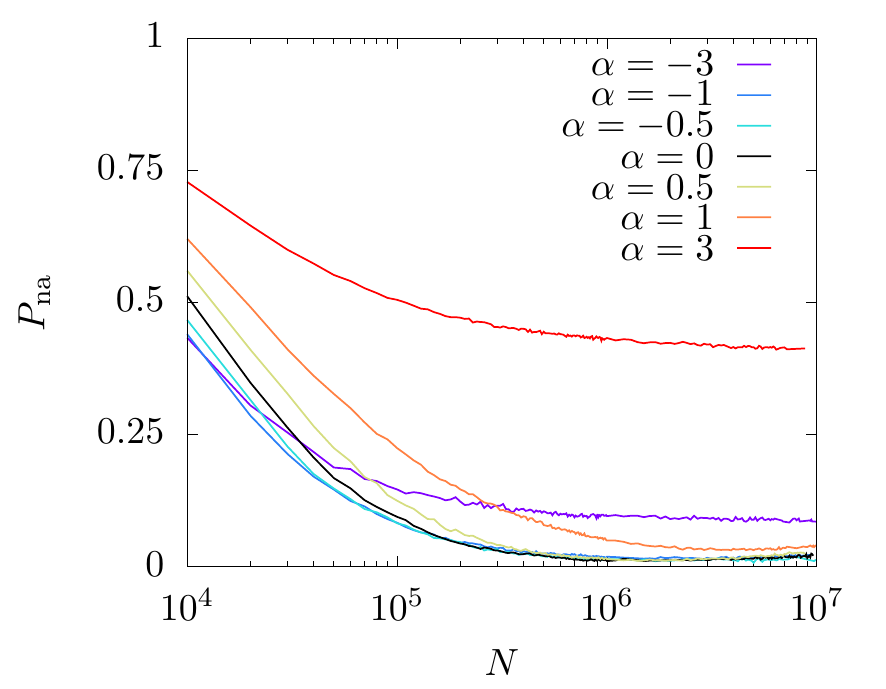}
    \caption{Simulation results of $P_\text{na}$ as a function of the detuning parameter $\alpha$ for two reciprocal values of the initial emittance ratio (top left), frequency excursion $\delta_\text{max}$ (top right), normal octupole strength $K_3$ (bottom left), and number of turns $N$ (bottom right). Seven values of $\alpha$ representing all possible regimes were used. A positive value of $\delta_\text{max}$ indicates a process where $\omega_{y,\text{i}}>\omega_x>\omega_{y,\text{f}}$ and vice-versa for $\delta_\text{max}<0$. (Simulation parameters: $\omega_x=2.602$, $\beta_x=\beta_y=1$, $\overline{\alpha_{xx}}=\overline{\alpha_{yy}}=0$, $\delta_\text{max}=0.01$, $K_3=10$, $N=10^6$, $N_p=10^4$, $\av{\Ii{x}} = \num{1e-4}$, $\av{\Ii{y}} =\num{4e-4}$).}
    \label{fig:axy_delta_k_n_pna} 
\end{figure*}

The quantity $h_z$ is the kurtosis of the beam distribution, which measures how the distribution reaches its peak in comparison with a Gaussian. Following~\cite{PhysRevSTAB.5.124202}, the standard normalization is modified so that $h_z=0$ for a Kapchinskij-Vladimirskii distribution~\cite{kv}, which is known to have no halo, and $h_z=1$ for a Gaussian distribution, so that $h_z>1$ indicates a halo larger than that of a Gaussian distribution.

Numerical simulations were used to study the behavior of $P_\text{na}$, $h_x$, and $h_y$ as a function of various parameters, namely the frequency excursion $\delta_\text{max}$, the octupole strength $K_3$, the ratio between initial emittances $\av{\Ii{y}} / \av{\Ii{x}}$, the number of turns $N$ of the resonance-crossing process, and the detuning parameter $\overline{\alpha_{xy}}$. For all numerical simulations, we set $\omega_x=2.602$, $\beta_x=\beta_y=1$, $\overline{\alpha_{xx}}=\overline{\alpha_{yy}}=0$. When not otherwise stated, we used the default parameters $\delta_\text{max}=0.01$, $K_3=10$, $N=10^6$, using $N_p=10^4$ initial conditions with $\av{\Ii{x}} = \num{1e-4}$, $\av{\Ii{y}} =\num{4e-4}$. Note that since $\delta(\lambda)$ in Eq.~\eqref{eq:ham} is defined as $\delta(\lambda) =\omega_x - \omega_y$, $\delta_\text{max}>0$ corresponds to a process in which $\delta$ (and $\eta$, if $K_3>0$) is varied from an initial negative value to a final positive value.

Figure~\ref{fig:axy_delta_k_n_pna} (top left) shows the performance of the emittance exchange as a function of the detuning parameter $\alpha$ for two reciprocal values of the initial emittance ratio. The values of $\alpha$ are obtained by varying the parameter $\overline{\alpha_{xy}}$ (the value corresponding to $K_3=10$ is shown on the top horizontal scale). It is clearly visible how $P_\text{na}$ is close to zero in the $-1<\alpha<1$ region, where the theory predicts a perfect exchange of emittance, while it grows considerably once $\vert \alpha  \vert > 1$. Furthermore, the behavior has a clear symmetry, \ie $P_\text{na}$ is invariant for transformations of the form
\begin{equation}
    \alpha\to -\alpha,\qquad \frac{\av{\Ii{x}}}{\av{\Ii{y}}} \to \left (\frac{\av{\Ii{x}}}{\av{\Ii{y}}} \right )^{-1} \, .
\end{equation}
Indeed, this transformation in actions corresponds to a transformation of type $J \to 1-J $ for the value of the variable $J=\Ja/\Jb=I_x/(I_x+I_y)$.

This can be explained by considering some further symmetries of the underlying Hamiltonian that are analyzed in the Appendix~\ref{app:symmetries}. 

\begin{figure}[htb]
    \centering
    \includegraphics[trim=1truemm 0truemm -2truemm 0truemm,width=0.54\columnwidth,clip=]{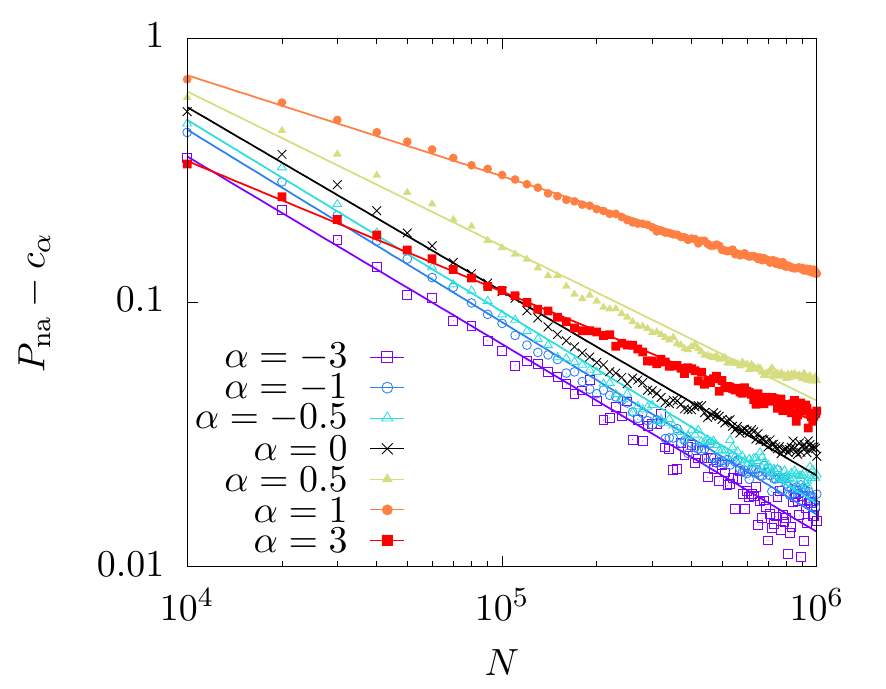}
    \includegraphics[width=0.54\columnwidth]{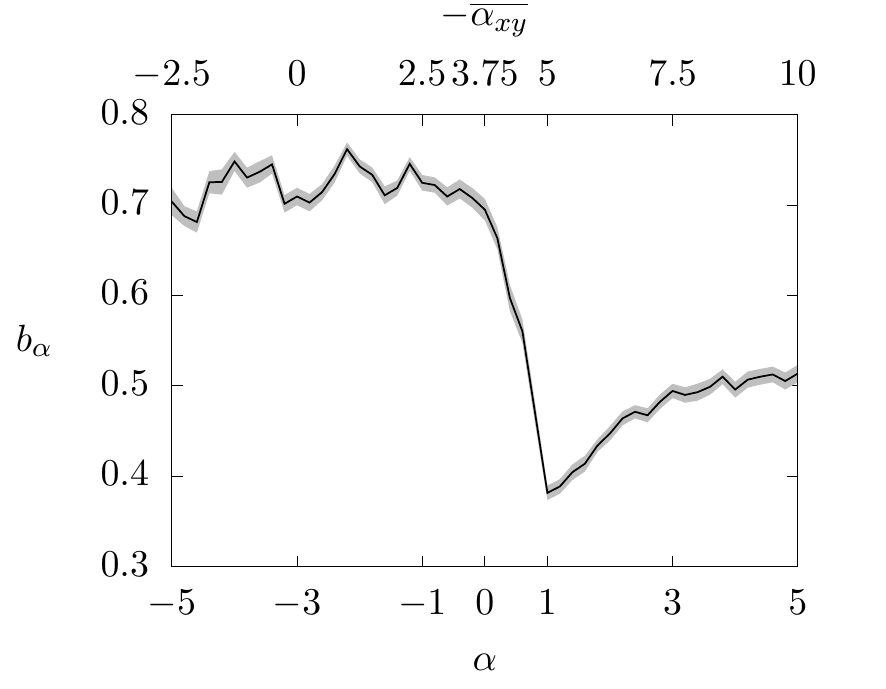}
    \caption{Top: log-log scale representation of the data of the bottom-right plot of Fig.~\ref{fig:axy_delta_k_n_pna}, having subtracted the offset $c_\alpha$ from $P_\text{na}$. The fit lines are also shown. Bottom: Values of the parameter $b_\alpha$ of the model $P_\text{na}(N) = a_\alpha N^{-b_\alpha} + c_\alpha$ that fits the data of the bottom-right plot of Fig.~\ref{fig:axy_delta_k_n_pna} as a function of the detuning coefficient $\alpha$, which is set by varying $\overline{\alpha_{xy}}$. The shaded area represents the error attributed to the computation of the fit parameter.}
    \label{fig:exp_vs_alpha}
\end{figure}


\begin{figure*}[htb]
    \centering
    \includegraphics[width=\textwidth]{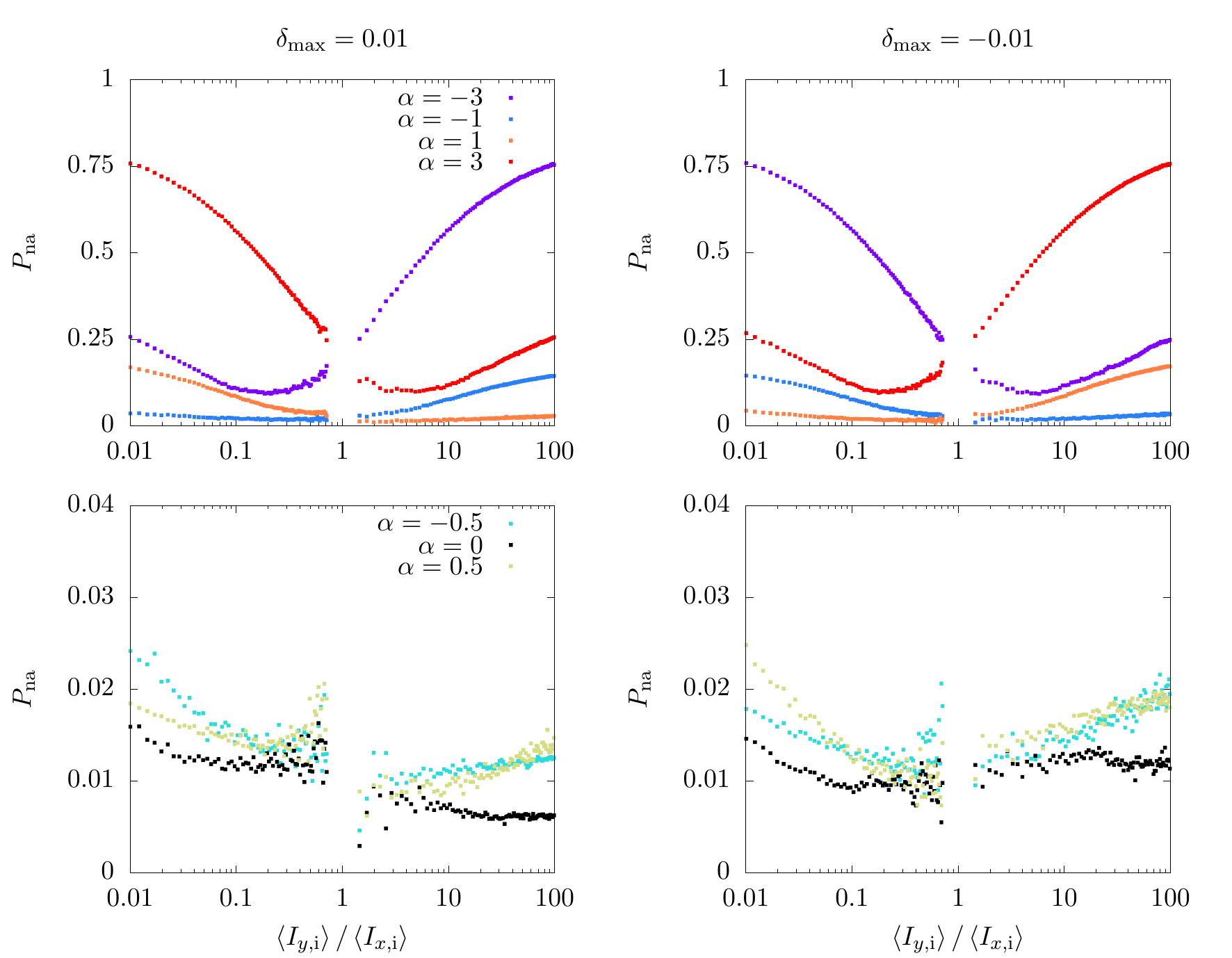}
    \caption{Simulation results of $P_\text{na}$ as a function of the initial emittance ratio $\av{\Ii{y}}/\av{\Ii{x}}$, for seven values of $\alpha$ representing all the possible regimes (divided into top and bottom plots) and two values of $\delta_\text{max}$ with different signs (left and right plots). Note the different vertical scale for plots with $\vert \alpha \vert \ge 1$ and $\vert \alpha \vert <1$. As $P_\text{na}$ diverges when $\av{\Ii{x}}\approx \av{\Ii{y}}$, data in the range $5/7 \le \av{\Ii{y}}/\av{\Ii{x}} \le 7/5$ have been excluded from the plot. (Simulation parameters: $\omega_x=2.602$, $\beta_x=\beta_y=1$, $\overline{\alpha_{xx}}=\overline{\alpha_{yy}}=0$, $K_3=10$, $N=10^6$, $N_p=10^4$, $\av{\Ii{x}} = \num{1e-4}$, $\av{\Ii{y}} =\num{4e-4}$).}
    \label{fig:sigma_pna}
\end{figure*}

In the other graphs of Fig.~\ref{fig:axy_delta_k_n_pna}, we present data sampled at seven values of $\overline{\alpha_{xy}}$, to test the exchange of emittance in the seven possible regimes of $\alpha$. Figure~\ref{fig:axy_delta_k_n_pna} (top right) shows $P_\text{na}$ as a function of the frequency excursion $\delta_\text{max}$, and in this case, $P_\text{na}$ fulfills the following symmetry
\begin{equation}
    \alpha\to -\alpha,\qquad \delta_\mathrm{max} \to -\delta_\mathrm{max} \, ,
\end{equation}
which is a direct consequence of the properties of the Hamiltonian expressed in Eq.~\eqref{eq:transformation}. 

Figure~\ref{fig:axy_delta_k_n_pna} (bottom left) shows the emittance exchange performance as a function of the octupole strength $K_3$, and in this case, the symmetry 
\begin{equation}
    \alpha\to -\alpha,\qquad K_3 \to -K_3 \, .
\end{equation}
is only approximately satisfied (this behavior is also discussed in Appendix~\ref{app:symmetries}).  

\begin{figure*}[htb]
    \centering
    \includegraphics[trim=0truemm 1truemm 0truemm 0truemm, width=\textwidth,clip=]{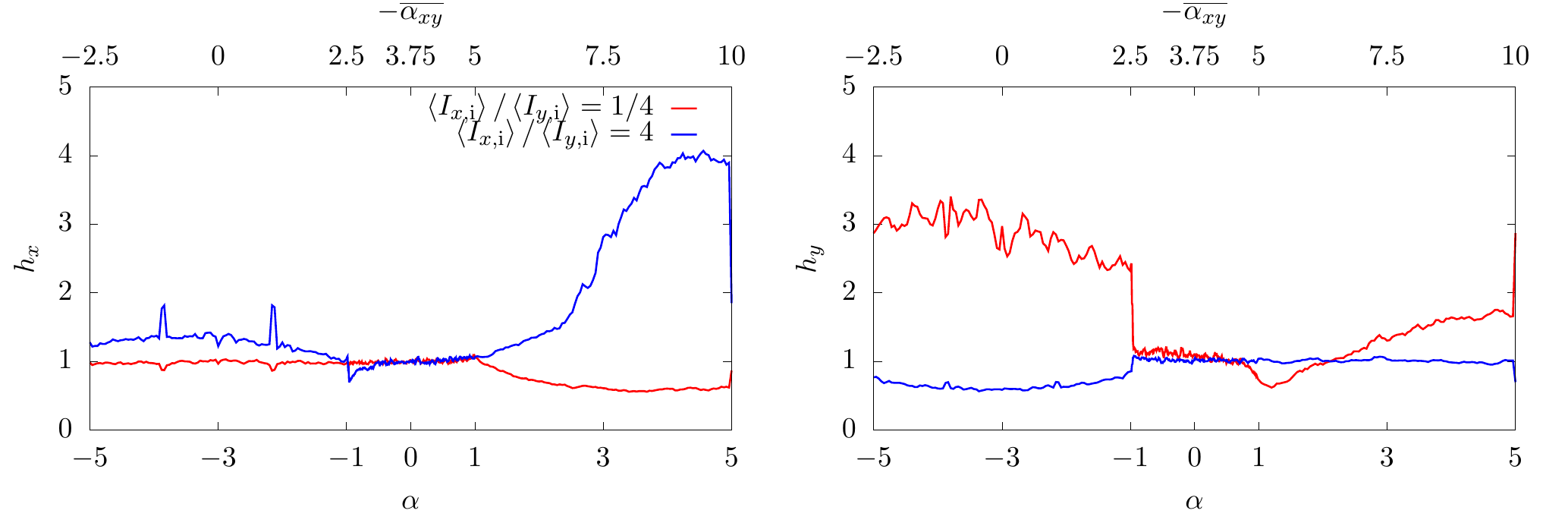}
    \includegraphics[trim=0truemm 1truemm 0truemm 2truemm, width=\textwidth,clip=]{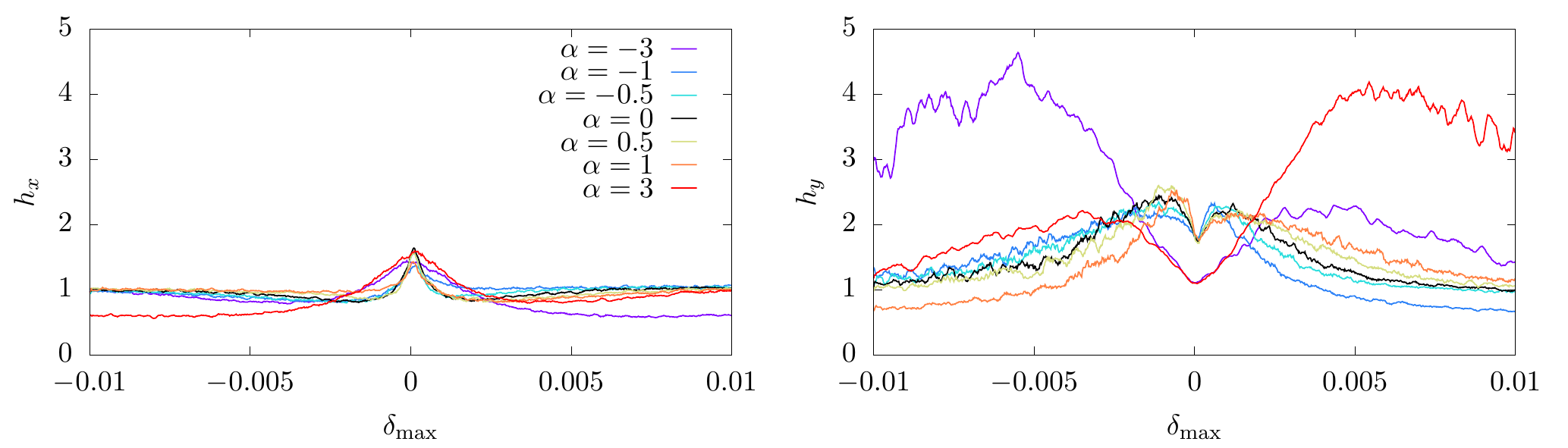}
    \includegraphics[trim=0truemm 1truemm 2.5truemm 2truemm, width=\textwidth,clip=]{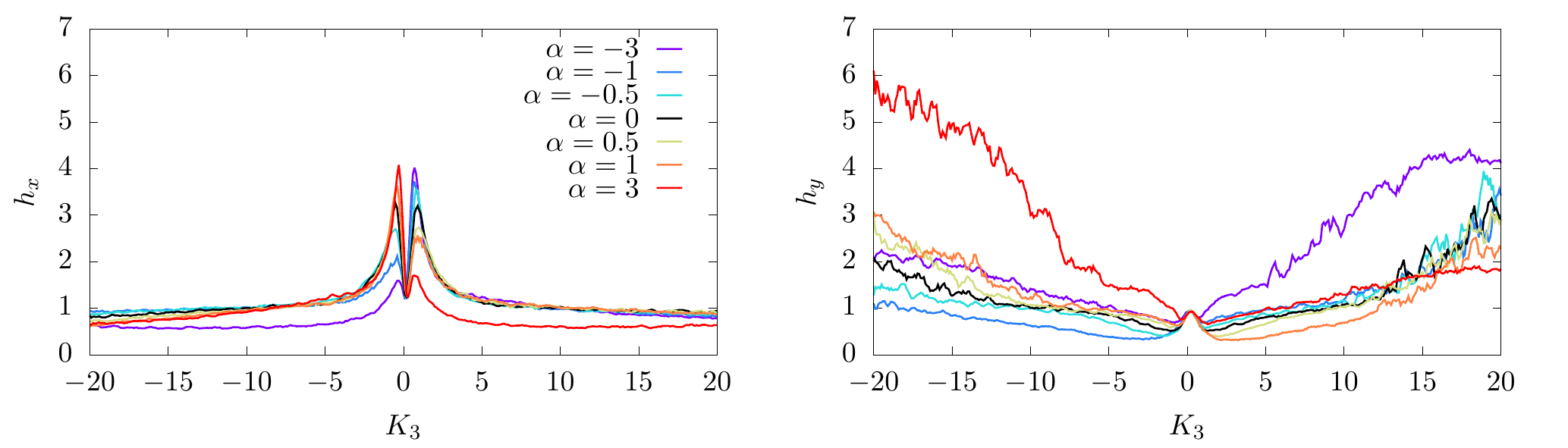}
    \includegraphics[trim=0truemm 1truemm 0truemm 2truemm, width=\textwidth,clip=]{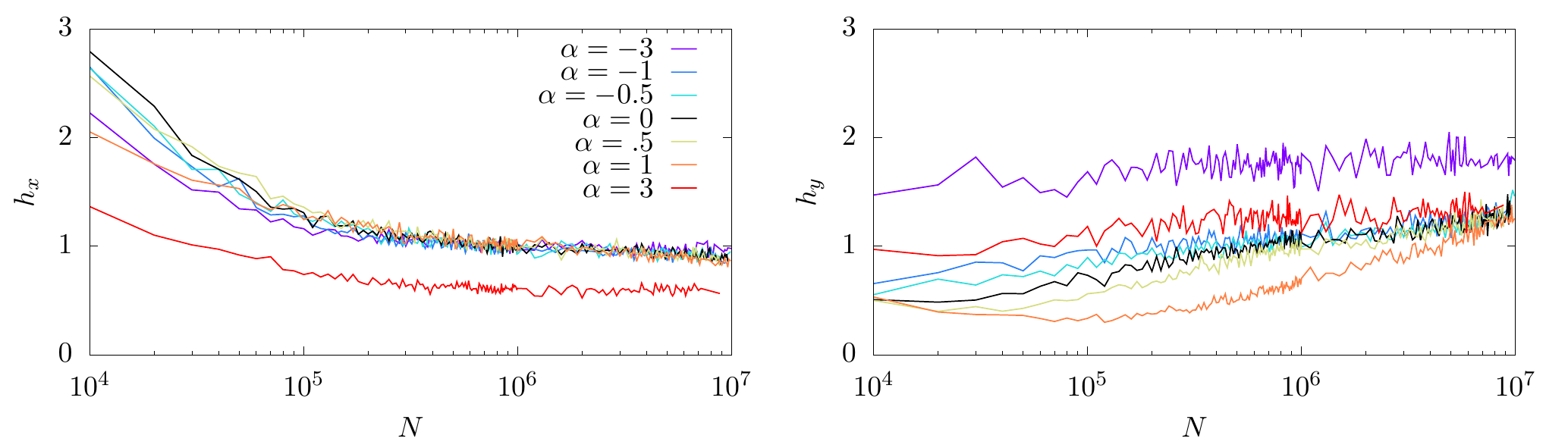}
    \caption{Simulation results of halo parameters $h_x$ (left) and $h_y$ (right) as a function of the detuning parameter $\alpha$ for two reciprocal values of the initial emittance ratio (first row), frequency excursion $\delta_\text{max}$ (second row), normal octupole strength $K_3$ (third row), and number of turns $N$ (fourth row). Seven values of $\alpha$ representing all possible regimes have been used. A positive value of $\delta_\text{max}$ indicates a process where $\omega_{y,\text{i}}>\omega_x>\omega_{y,\text{f}}$ and vice-versa for $\delta_\text{max}<0$. Data are shown after applying a moving average of $5$ values. (Simulation parameters: $\omega_x=2.602$, $\beta_x=\beta_y=1$, $\overline{\alpha_{xx}}=\overline{\alpha_{yy}}=0$, $\delta_\text{max}=0.01$, $K_3=10$, $N=10^6$, $N_p=10^4$, $\av{\Ii{x}} = \num{1e-4}$, $\av{\Ii{y}} =\num{4e-4}$).}
    \label{fig:axy_delta_k_n_halo}
\end{figure*}



Figure~\ref{fig:axy_delta_k_n_pna} (bottom right) shows $P_\text{na}$ as a function of the number of turns used for the map simulation, which corresponds to the study of the efficiency of emittance exchange as a function of the inverse of the adiabaticity parameter. As expected, the results are well fitted by a power-law model with an offset, \ie $P_\text{na} = a_\alpha N^{-b_\alpha} + c_\alpha$. This is the typical behavior when separatrices are present in phase space (see Ref.~\cite{PhysRevAccelBeams.24.094002} and references therein). In Fig.~\ref{fig:exp_vs_alpha} (top), $P_\mathrm{na}$ is shown in log-log scale as a function of $N$ after subtraction of the constant term $c_\alpha$. The fitted straight lines are also shown and the good agreement is clearly visible, which confirms the statement about the type of model that best represents the dependence of $P_\mathrm{na}$ on $N$. In Fig.~\ref{fig:exp_vs_alpha} (bottom), the value of the fit parameter $b_\alpha$ as a function of $\alpha$ is shown, and the shaded area represents the error associated with the computation of the fit parameter. Two regimes are visible: one corresponding to the case $\vert \alpha \vert \leq 1$, when the emittance exchange occurs; one corresponding to the case $\vert \alpha \vert > 1$, when only partial emittance exchange occurs, due to the bifurcation phenomenon of the origin. It should be stressed, however, that the details of the observed functional dependence of these parameters on $N$ and $\alpha$ is model-dependent. 

Figure~\ref{fig:sigma_pna} reports the value of $P_\text{na}$ as a function of the initial emittance ratio, for the two directions of resonance crossing, \ie $\delta_\text{max}>0$ (left column) and $\delta_\text{max}<0$ (right column), and the seven values of $\alpha$ already considered (the largest in the top row and those closer to zero in the bottom row). These plots condense the behavior already shown in the previous figure. Note the symmetry between the logarithm of the emittance ratio, the sign of $\alpha$ and the sign of $\delta_\text{max}$. Furthermore, it should be noted that the vertical scales are different in the bottom plots, as $P_\text{na}\approx 0$ when $\alpha<1$. It is also worth noting that for $\langle \Ii{x} \rangle = \langle \Ii{y} \rangle$, $P_\text{na}$ (Eq.~\ref{eq:pna}) is not defined, which explains the fluctuations of $P_\text{na}$ as the ratio approaches $1$.

Figure~\ref{fig:axy_delta_k_n_halo} shows the dependence of the halo parameter in the horizontal (left column) and vertical (right column) planes as a function of various parameters (reported in the rows) that characterize the model.The data shown represent a moving average as the values of the halo parameters are rather noisy due to the sensitivity to the presence of particles in the tails of the distribution with the fourth power in the definition of $h_z$. Isolated outliers found at the extremities of the initial distribution are therefore responsible for the very large values of $h_z$ found in the numerical simulations.

In general, the symmetries observed for $P_\text{na}$ are also very visible for the halo parameters. The vertical plane features larger values of the halo parameter, which is likely to be  linked to the choice of the distribution of the initial conditions, which satisfy $\langle \Ii{x} \rangle < \langle \Ii{y} \rangle$. In fact, a similar situation would be found in the horizontal plane upon reversing the shape of the distribution of the initial conditions, \ie having $\langle \Ii{x} \rangle > \langle \Ii{y} \rangle$. When a full emittance exchange is expected, \eg when $\alpha < 1$ or $\left \vert \delta_\text{max}\right \vert $ large, and hence $P_\text{na}\approx 0$, both $h_x$ and $h_y$ approach $1$, confirming that the final distribution is still Gaussian. The dependence of $h_z$ on the number of turns $N$ shows a convergence towards $1$ in the horizontal plane. This indicates that a slower, and hence more adiabatic, resonance crossing is beneficial not only for a good exchange of the transverse emittances, but also for ensuring that the final distribution is still Gaussian, as expected from the estimates in~\cite{NEISHTADT1986}. The situation in the vertical plane is somewhat different, as a small increase of $h_y$ is observed as a function of $N$. Note that particles with final $x$ or $y$ greater than $5$ times the value of the standard deviation have been filtered out. Even so, the small number of outlier particles tend to move further away from the origin with each iteration of the map, which explains the increasing trend of $h_y$. 
\begin{figure*}[htb]
    \centering
    \includegraphics[trim=0truemm 0truemm 0truemm 4truemm, width=\textwidth,clip=]{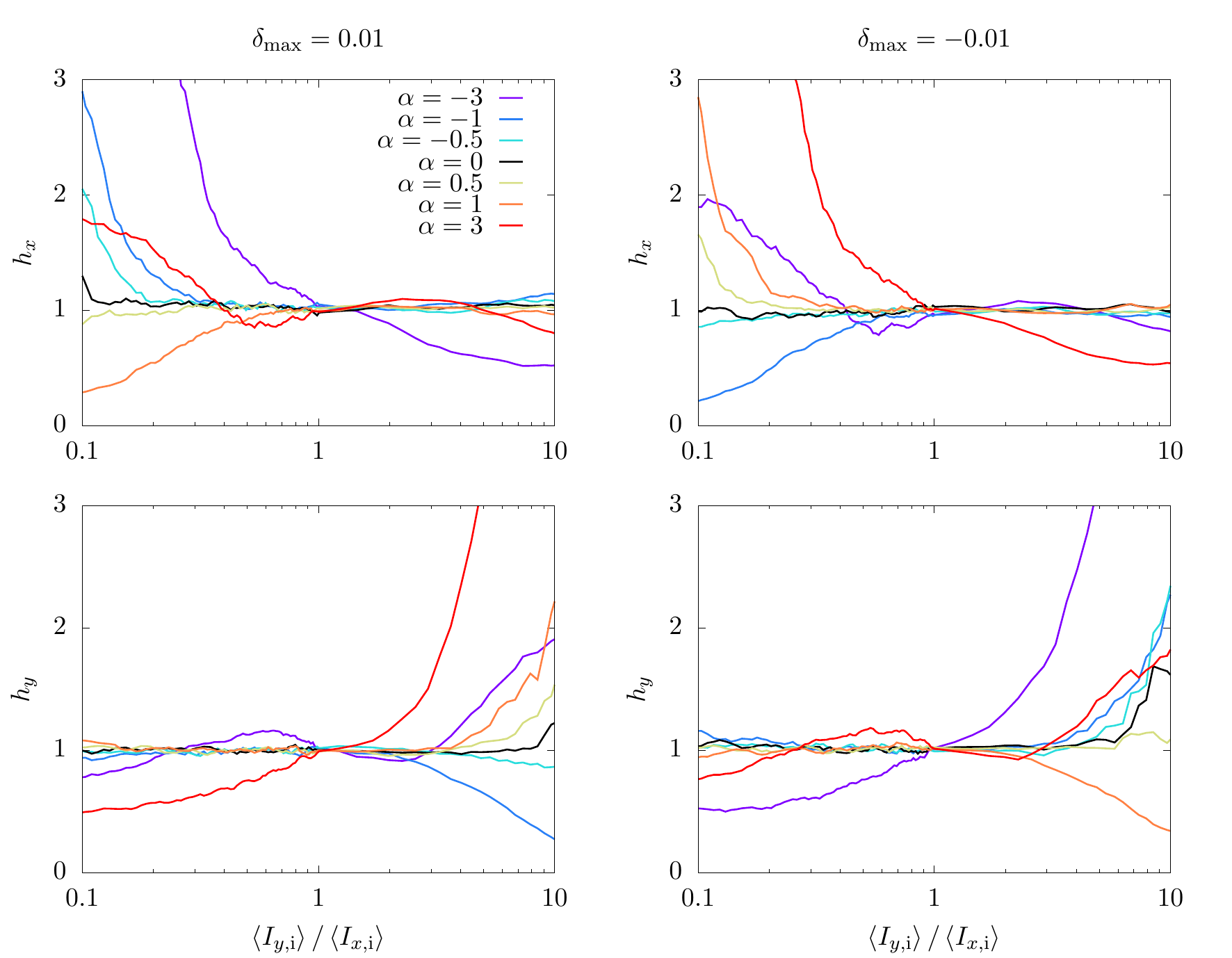}
    \caption{Simulation results of the halo parameters $h_x$ (top plots) and $h_y$ (bottom) as a function of the initial emittance ratio $\av{\Ii{y}}/\av{\Ii{x}}$, for seven values of $\alpha$ representing all possible regimes and two values of $\delta_\text{max}$ with different signs (left and right). Data are shown after applying a moving average of $5$ values. (Simulation parameters: $\omega_x=2.602$, $\beta_x=\beta_y=1$, $\overline{\alpha_{xx}}=\overline{\alpha_{yy}}=0$, $\delta_\text{max}=0.01$, $K_3=10$, $N=10^6$, $N_p=10^4$, $\av{\Ii{x}} = \num{1e-4}$, $\av{\Ii{y}} =\num{4e-4}$).}
    \label{fig:sigma_halo}
\end{figure*}

Finally, in Fig.~\ref{fig:sigma_halo}, we show the halo parameters $h_x$ (top row) and $h_y$ (bottom row) for opposite values of $\delta_\text{max}$ (left and right columns) and for the seven values of $\alpha$, as a function of $\av{\Ii{x}} / \av{\Ii{y}}$. It is clearly seen that for $\vert \alpha \vert <1$ the final distribution is Gaussian-like over a wide range of values of $\av{\Ii{x}} / \av{\Ii{y}}$, while the final distribution deviates from a Gaussian in the regimes $|\alpha|>1$. A symmetry between the conditions with $\av{\Ii{y}}>\av{\Ii{x}}$ or $\av{\Ii{y}}<\av{\Ii{x}}$, the direction of the resonance-crossing process, represented by the sign of $\delta_\text{max}$ and the sign of $\alpha$ is also visible.
\section{Conclusions} \label{sec:conc}
A Hamiltonian model to describe the crossing of the nonlinear coupling resonance, known as the Montague resonance whenever space charge forces are at its origin, has been proposed and analyzed in detail. The phase-space topology has been studied in combination with the process of adiabatic trapping induced by the time variation of the area of the different phase-space regions that appear when the resonance is crossed. 

These analyses show that exchange of the transverse emittances is indeed possible and that the actual performance depends on the adiabaticity of the crossing process and on the detail of the phase-space topology generated during the resonance crossing process.

By combining the recent results on emittance exchange for the linear coupling resonance with the manipulation of transverse emittances by crossing nonlinear 2D resonances described in this paper, it can be concluded that for the Montague resonance, which induces a nonlinear coupling of the transverse planes, the efficiency of emittance exchange is not always granted, as the presence of separatrices may prevent this. This is a key difference with respect to the crossing of the linear coupling resonance, where emittance exchange always occurs. Another essential difference is that the performance of the emittance exchange process with the number of turns used to cross the resonance is represented by a power law, while in the case of the linear coupling resonance this dependence is exponential. These differences stem from the singularity of the action-angle variables for the frozen Hamiltonian at the separatrix curves that have been shown to exist in phase space for certain values of the model parameters. No separatrix is present in the linear case.

The impact of the nonlinearities on the behavior of the Montague resonance makes it important to study the beam halo during the resonance-crossing process. It has been shown that for certain initial conditions, the resonance crossing process may induce large fluctuations in the adiabatic invariant to produce a beam halo in the final distribution. Detailed numerical simulations have revealed that the halo parameter, introduced to study these fluctuations, inherits the symmetries of the Hamiltonian system with halo formation mitigated by improving the adiabaticity of the resonance crossing.  

The mechanisms discussed here neglect the presence of periodic variations of the transverse tunes that may be induced, \eg by a non-zero value of the linear chromaticity. These effects might induce trapping and detrapping phenomena that could have a negative impact on the performance of the emittance exchange process. The situation presented can therefore be considered as a best-case scenario of what could occur in a real machine. A mitigation for this effect would be the reduction of the linear chromaticity towards zero, provided this can be achieved ensuring beam stability. The approach proposed in this paper has also been shown to be applicable when there is a time dependence of the resonance strength, which is relevant when space-charge effects are considered.

The understanding of the details of emittance exchange and halo formation generated by the crossing of the nonlinear coupling resonance presented in this paper may be of great benefit to mitigate the harmful effects induced by an uncontrolled crossing of the Montague resonance, which is a rather common occurrence during the operation of circular accelerators.
\section*{Acknowledgments}
We would like to thank Y.~Papaphilippou for very useful discussions on the original version of this manuscript. 
\clearpage
\appendix
\section{Derivation of the Hamiltonian model} \label{app:derivation}
The linear dynamics of a Poincar\'e section of motion of a charged particle in a circular accelerator is described by a symplectic map that, without loss of generality, corresponds to the phase flow of a harmonic oscillator with phase advance $\omega$, where $\omega L$ is the linear tune for a circumference length $L$. The presence of an adiabatic variation in the quadrupole fields changes the tune as $\omega(\epsilon N)$ where $N$ is the number of turns. In such a case it is possible to interpolate the linear dynamics using a time-dependent quadratic Hamiltonian 
\begin{equation}
    \begin{split}
    H({p}_x,{p}_y,{x},{y},\lambda) & = \frac{{p}_x^2+{p}_y^2}{2} + \\
    & + \frac{1}{2} \left ( \omega_x^2(\lambda)\,  x^2+\omega_y^2(\lambda) \, y^2 \right ) \, ,
    \end{split}
  \label{eq:hamqlin}
\end{equation}
where $\lambda=\lambda(s)$ with $\dd \lambda/\dd s=O(\epsilon)$, $\epsilon$ being the adiabatic parameter, and we use physical coordinates (the momenta are normalized with respect to the total momentum). The presence of a pseudo-octupolar term, whose strength is represented by the coefficient $G$, can be introduced to selectively excite the resonance $(2,-2)$, and the new Hamiltonian reads
\begin{equation}
    \begin{split}
    H({p}_x,{p}_y,{x},{y},\lambda) & = \frac{{p}_x^2+{p}_y^2}{2} + \\
    & + \frac{1}{2} \left ( \omega_x^2(\lambda)\,  {x}^2+\omega_y^2(\lambda) \, {y}^2 + 2 G \, {x}^2 {y}^2 \right ) \, ,
    \end{split}
  \label{eq:hamq}
\end{equation}
where we assume $G$ constant. By varying $\lambda$, we describe the resonance crossing process in the adiabatic approximation, \ie $\epsilon\to 0$. The results of adiabatic theory do not depend on the explicit form of the function $\lambda(\epsilon)$, so one can assume $\lambda=\epsilon s$ also in the Hamiltonian~\eqref{eq:hamq} without loss of generality. In this case the linear normal form introduces a further term in the original Hamiltonian. If we indicate with $\bm{A}(\lambda)$ the matrix of the transformation $Z =z \, \sqrt{\omega_z(\lambda)}$, it induces the transformation  
\begin{equation}
\bm x=\bm A(\lambda) \, \bm X\, ,
\end{equation}
where $\bm X$ are the new coordinates. A generating function $F_2( \bm x,\bm P, \lambda)$ for the symplectic transformation can be written in the form
\begin{equation}
F_2(\bm x,\bm P, \lambda)=\bm P^\top \, \bm A^{-1}(\lambda) \, \bm x 
\end{equation}
and the new Hamiltonian reads
\begin{equation}
\begin{split}
H(\bm X,\bm P,\lambda) &=\omega_x(\lambda)\frac{X^2+ P_x^2}{2}+\omega_y(\lambda)\frac{Y^2+P_y^2}{2}+\\
& +\frac{G}{{\omega_x(\lambda) \omega_y(\lambda)}} X^2\, Y^2 +\epsilon \, \bm P^\top \pdv{\, \bm{A}^{-1}}{\lambda} \bm A \bm X \, ,
\end{split}
\end{equation}
where the last term is the time derivative of the generating function. The final form of the Hamiltonian is as follows
\begin{equation}
    \begin{split}
H(\bm X, \bm P, \lambda) & =\omega_x(\lambda)\frac{X^2+P_x^2}{2}+\omega_y(\lambda) \frac{Y^2+P_y^2}{2}+\\
& + \frac{G}{{\omega_x(\lambda)\omega_y(\lambda)}} X^2 \, Y^2 + \\
& + \frac{\epsilon}{2}\left [\frac{\omega_x'(\lambda)}{\omega_x(\lambda)}X\, P_x + \frac{\omega_y'(\lambda)}{\omega_y(\lambda)}Y \, P_y\right ] \, ,
    \end{split}
    \label{eq:hamt}
\end{equation}
where $\omega'=\dd\omega/\dd\lambda$. The linear action angle variables $(\bm \theta, \, \bm I)$ can be used to recast the Hamiltonian~\eqref{eq:hamt} in the form
\begin{equation}
    \begin{split}
H(\bm \theta,\bm I, \lambda)&= \omega_x(\lambda) I_x+\omega_y(\lambda) I_y+
\frac{2 G}{{\omega_x(\lambda)\omega_y(\lambda)}} \times \\
& \times {I_x I_y}\sin^2\theta_x\sin^2\theta_y + \\
 + \epsilon \left [\frac{\omega_x'(\lambda)}{\omega_x(\lambda)} \right .&  \left . I_x\sin\theta_x\cos\theta_x+  \frac{\omega_y'(\lambda)}{\omega_y(\lambda)}I_y\sin\theta_y\cos\theta_y\right ] \, .
    \end{split}
    \label{eq:hamapp1}
\end{equation}

The introduction of a slow phase $\phia=\theta_x-\theta_y$ in the generating function
\begin{equation}
    F_2(\bm \theta, \bm J)=\begin{pmatrix}
\Ja, & \Jb
\end{pmatrix}
\begin{pmatrix}
1 &          - 1 \\
0 & \phantom{-}1
\end{pmatrix}
\begin{pmatrix}
\theta_x\\ \theta_y
\end{pmatrix}
\label{eq:genfun}
\end{equation}
transforms the Hamiltonian into the form
\begin{equation}
    \begin{split}
        H(\bm \phi, \bm J,\lambda) & =\delta(\lambda) \Ja+\omega_y \Jb+
\frac{2 G}{{\omega_x(\lambda)\omega_y(\lambda)}} \times \\
& \times {\Ja (\Jb-\Ja)}\sin^2(\phia+\phib)\sin^2\phib+ \\
&+
\epsilon\left [\frac{\omega_x'(\lambda)}{\omega_x(\lambda)}\Ja\sin(\phia+\phib)\cos(\phia+\phib)+ \right . \\
& + \left . \frac{\omega_y'(\lambda)}{\omega_y(\lambda)}(\Jb-\Ja)\sin\phib\cos\phib\right ]
    \end{split}
\end{equation}
where
\begin{equation}
    \delta(\lambda)=\omega_x(\lambda)-\omega_y(\lambda)
\end{equation}
and it is possible to apply a perturbative approach averaging over the fast-evolving angle $\phib$ to obtain the Hamiltonian
\begin{equation}
    \begin{split}
        H(\bm \phi, \bm J,\lambda) & =\delta(\lambda) \Ja+\omega_y \Jb+
\frac{\pi G}{{2 \omega_x(\lambda)\omega_y(\lambda)}} \times \\
& \times {\Ja (\Jb-\Ja)}\cos2 \phia+
O(\epsilon^2) \, .
    \end{split} 
    \label{eq:for_referee}
\end{equation}

It is readily seen that the coefficient multiplying the resonant term is indeed time-independent and, with a slight abuse of notation, will be renamed $G$ and all other terms dropped.  

As $\phi_\mathrm{b}$ is not present in the Hamiltonian, it follows that $\Jb$ is constant up to an error $O(\epsilon^2)$ for a time interval of order $O(\epsilon^{-1})$. The perturbative approach is possible only if this error is small, so that $\Jb$ can be considered constant during the resonance crossing process. We remark that the term $\omega_y \Jb$ can be dropped as it affects only the dynamics of $\phi_\mathrm{b}$, which is irrelevant in the case under consideration. In such a case, the action of the 1DoF Hamiltonian
\begin{equation}
    H(\bm \phi,\bm J,\lambda) = \delta(\lambda) \Ja + G {\Ja (\Jb-\Ja)}\cos 2 \phia \, .
\label{hamsphe}
\end{equation}
can be considered an adiabatic invariant up to an error $O(\epsilon^{-1})$ for a time interval $O(\epsilon^{-1})$, and we can study the change of $\Ja$ when $\delta(\lambda)$ passes through zero. 

An amplitude detuning term of the form $ \alpha_{xx}I_x^2 + 2 \alpha_{xy}I_x I_y + \alpha_{yy} I_y^2$ can also be included in the model, and following the same steps one arrives at the following model 
\begin{equation}
\begin{split}
    \ham (\phia, & \Ja, \Jb) = \delta (\lambda) \Ja + \frac{1}{2} \alfaa \Ja^2 + \alfab \Ja \Jb + \\
    + & G \Ja (\Jb-\Ja) \cos 2\phia + \qty[\omega_y(\lambda) \Jb + \frac{1}{2}\alfbb\Jb^2] \, ,
\end{split}
\label{eq:ham12}
\end{equation}
where 
\begin{equation}
\begin{split}
   \frac{1}{2} \alfaa & = \alpha_{xx}-2\alpha_{xy}+\alpha_{yy},\\
   \alfab             & =2\alpha_{xy}-2\alpha_{yy},\\
   \frac{1}{2} \alfbb & = \alpha_{yy} \, ,
\end{split}
\label{eq:ampdet1}
\end{equation}
and the term in the square brackets in Eq.~\eqref{eq:ham12} can be dropped as the Hamiltonian does not depend on $\phib$.
\section{Normal Form Hamiltonian} \label{app:nf}
The relationship between the Hamiltonian of Eq.~\eqref{eq:hamf} and the map of Eq.~\eqref{eq:henon} can be obtained by using Normal Form procedure outlined in~\cite{Bazzani:262179} to calculate the resonant interpolating Hamiltonian for the 4D H\'enon-like map in presence of a normal octupole, considering also the extra detuning terms of Eq.~\eqref{eq:detuning}.

First of all, Eq.~\eqref{eq:henon} can be written introducing the complex coordinates $z_1=x-ip_x$, $z_2=y-ip_y$ ($z^\ast$ indicating the complex conjugate of $z$) as $\bm z' = \bm F(\bm z)$, where
\begin{equation}
\begin{aligned}
    \bm F_1 &= \exp(i\omega_x + i\overline{\alpha_{xx}} z_1 z_1^\ast/2 +  i\overline{\alpha_{xy}} z_2 z_2^\ast/2) \times \\ 
    & \qquad \begin{aligned}  \Bigg\{ z_1 &- i\frac{K_3}{48}\left [\beta_x^2 (z_1+z_1^\ast)^3+\right. \\
    &  \left.  - 3\beta_x\beta_y(z_1+z_1^\ast)  (z_2+z_2^\ast)^2 \right ] \Bigg \}\,, \end{aligned} \\
    \bm F_2 &= \exp(i\omega_y + i\overline{\alpha_{xy}} z_1 z_1^\ast/2 +  i\overline{\alpha_{yy}} z_2 z_2^\ast/2)\times \\ &\qquad \begin{aligned} \Bigg \{ z_2 &- i\frac{K_3}{48}\left [\beta_y^2 (z_2+z_2^\ast)^3+ \right. \\ &\left. - 3\beta_x\beta_y(z_1+z_1^\ast)^2(z_2+z_2^\ast) \right ] \Bigg \}\,.\end{aligned}\\
\end{aligned}
\end{equation}

As we are interested in the resonant Normal Form $\bm U$, i.e. the solution of the homological equation 
\begin{equation}
 \bm{F}\circ \bm{\Phi} = \bm \Phi \circ \bm U 
 \label{eq:homological}
\end{equation}
where $\bm \Phi$ is a coordinate transformation, for the coupling resonance, we can set $\omega_x=\omega_y$. Up to order $3$, $\bm U(\bm \zeta, \bm \zeta^\ast)$ is given by
\begin{equation}
\begin{split}
   \bm U_j =e^{i  \omega_{x}}\zeta_j &+ u_{j,{2100}} {\zeta_1}^{2} {\zeta_1^\ast} + u_{j,{1110}} {\zeta_1}
{\zeta_1^\ast} {\zeta_2} + u_{j,{0120}} {\zeta_1^\ast} {\zeta_2}^{2} \\ &+
u_{j,{2001}} {\zeta_1}^{2} {\zeta_2^\ast} + u_{j,{1011}} {\zeta_1}
{\zeta_2} {\zeta_2^\ast} + u_{j,{0021}} {\zeta_2}^{2} {\zeta_2^\ast} \, ,
\end{split}
\end{equation}
with $j=1,2$, and where only the resonant monomials have been considered.

There are no coefficients of $\bm U$ at order $2$, therefore the conjugating function $\Phi$ is the identity up to order $2$, and we should solve Eq.~\eqref{eq:homological} starting from order $3$. Comparing term by term the polynomials on the two sides of the homological equation, one obtains the following result
\begin{equation}
\begin{aligned}
    u_{1,0021}&=0, & u_{1,0120}&=i\frac{K_3}{16}\beta_x\beta_ye^{i\omega_x}, \\
    u_{1,1110}&=0, & u_{1,1011}&=i\qty(\frac{K_3}{8}\beta_x\beta_y + \frac{1}{2}\overline{\alpha_{xy}})e^{i\omega_x}, \\
    u_{1,2001}&=0, & u_{1,2100}&=-i\qty(\frac{K_3}{16}\beta^2_x - \frac{1}{2}\overline{\alpha_{xx}})e^{i\omega_x}, \\
    u_{2,0120}&=0, & u_{2,0021}&=-i\qty(\frac{K_3}{16}\beta^2_y - \frac{1}{2}\overline{\alpha_{yy}})e^{i\omega_x}, \\
    u_{2,1011}&=0, & u_{2,1110}&=i\qty(\frac{K_3}{8}\beta_x\beta_y + \frac{1}{2}\overline{\alpha_{xy}})e^{i\omega_x}, \\
    u_{2,2100}&=0, & u_{2,2001}&=i\frac{K_3}{16}\beta_x\beta_ye^{i\omega_x}. \\
\end{aligned}
\end{equation}

The interpolating resonant Hamiltonian $\ham_\text{res} = -iH$ can be computed using the Lie operator method. Defining recursively the Lie derivative as %
\begin{equation}
    D^0_H\bm \zeta = \bm \zeta, \qquad D^j_H \bm \zeta = \{ D^{j-1}_H \bm\zeta , H \}\qquad j>0
\end{equation}
where $\{\cdot,\cdot\}$ stands for the Poisson bracket, we can find that the interpolating Hamiltonian starts at order $4$, and
\begin{equation}
    \pdv{[H]_4}{\zeta^\ast_j} = e^{-i\omega_x}[U_j]_3 - \qty[ \sum_{k=0}^2 \frac{1}{k!} D^k_{[H]_{\le 3}}\zeta_j]_3 = h'_j(\bm \zeta)
    \label{eq:interHam}
\end{equation}
with $j=1,2$, where the symbol $[\cdot]_n$ represents the terms of order $n$ of a given polynomial.

An integral of Eq.~\eqref{eq:interHam} can be found in the form $H= h'_1(\bm\zeta)\,\zeta_1^\ast /2 + h'_2(\bm\zeta)\,\zeta_2^\ast /2$, and we obtain the Hamiltonian
\begin{equation}
\begin{split}
    \ham_\text{res} = &\qty(\frac{\overline{
\alpha_{xx}}}{4} -\frac{K_3}{32} \beta_{x}^{2}) {\zeta_1}^{2} {\zeta_1^\ast}^{2} + \\
           + & \qty(\frac{K_3}{8} \beta_{x} \beta_{y} + \frac{\overline{\alpha_{xy}}}{2}){\zeta_1} {\zeta_1^\ast} \zeta_2 \zeta_2^\ast + \\
        + &  \qty(- \frac{K_3}{32} \beta_{y}^{2} + \frac{\overline{\alpha_{yy}}}{4} ){\zeta_2}^{2} {\zeta_2^\ast}^{2} + \\
           + &  \frac{K_3}{32} \beta_{x} \beta_{y} \qty( {\zeta_1^\ast}^{2}
\zeta_2^2 + \zeta_1^{2} {\zeta_2^\ast}^{2}) \, . 
\end{split}
\end{equation}

As the transformation $\bm \Phi$ is the identity up to order $2$, in the Hamiltonian we can replace $\bm \zeta$ with $\bm z$, and using the action angle coordinates $z_1 = \sqrt{2I_x}e^{i\phi_x}$, $z_2=\sqrt{2I_y}e^{i\phi_y}$, one finds
\begin{equation}
\begin{split}
    \ham_\text{res} = & \qty(\overline{\alpha_{xx}} - \frac{K_3}{8}\beta_{x}^{2})  I_{x}^{2} + \\
    + & \qty(\frac{K_3}{2} \beta_{x} \beta_{y} + 2 \overline{\alpha_{xy}}) I_x I_y + \\
 + & \qty(\overline{\alpha_{yy}} - \frac{K_3}{8}\beta_{y}^{2})  I_{y}^{2} + \\
    + & \frac{K_3}{4} \beta_{x} \beta_{y} I_{x} I_{y}  \cos 2(\phi_x - \phi_y)\,.
 \end{split}
\end{equation}

If $\omega_x=\omega_y+\delta$, the quasiresonant Hamiltonian $\ham$, in leading order in $\delta$, is given by $\omega_x I_x + \omega_y I_y + \ham_\text{res}$, and we can perform the transformation to the coordinates $(\phi_\mathrm{a},\,J_\mathrm{a})$ as in Eq.~\eqref{eq:ham12}, which yields
\begin{equation}
    \begin{split}
        \ham &= \delta\Ja + \frac{K_3}{4}\beta_x\beta_y \Ja(\Jb-\Ja)\cos 2\phia\\
        &+ \left (\overline{\alpha_{xx}}-2\overline{\alpha_{xy}}+\overline{\alpha_{yy}}- \frac{K_3}{8}\beta^2_x - \frac{K_3}{2}\beta_x\beta_y - \frac{K_3}{8}\beta_y^2 \right ) \Ja^2 \\
        &+ \qty( \frac{K_3}{2}\beta_x\beta_y + \frac{K_3}{4}\beta_y^2 + 2\overline{\alpha_{xy}}-2\overline{\alpha_{yy}})\Ja\Jb\,.
    \end{split}
\end{equation}
From this, the correspondence between the quantities $G$, $\alfaa$ and $\alfab$ of Eq.~\ref{eq:ham} and the map parameters $K_3$, $\beta_x$, $\beta_y$, $\overline{\alpha_{xx}}$, $\overline{\alpha_{xy}}$, $\overline{\alpha_{yy}}$ can be established. 
\section{Transformation rules of the resonant Hamiltonian} \label{app:symmetries}
The Hamiltonian~\eqref{eq:hamf} changes sign under the transformation~\eqref{eq:transformation}, leaving the equations of motion invariant. However, it fulfills also other symmetries that should be studied to interpret the results shown in Fig.~\ref{fig:axy_delta_k_n_pna}. Starting from the top-left plot, $P_\text{na}$ is invariant under the transformation
\begin{equation}
    \alpha\to -\alpha,\qquad \frac{\av{\Ii{x}}}{\av{\Ii{y}}} \to \left (\frac{\av{\Ii{x}}}{\av{\Ii{y}}} \right )^{-1} \, ,
\end{equation}
and this can be studied by considering the exchange of the two transverse planes, \ie $I_x \leftrightarrow I_y$ and $\theta_x \leftrightarrow \theta_y$ in the Hamiltonian~\eqref{eq:hamapp1}, which can be recast in the form
\begin{equation}
    \begin{split}
H_{I_x \leftrightarrow I_y}(\bm \theta,\bm I, \lambda)&= \omega_x(\lambda) I_y+\omega_y(\lambda) I_x+
\frac{2 G}{{\omega_x(\lambda)\omega_y(\lambda)}} \times \\
& \times {I_x I_y}\sin^2\theta_x\sin^2\theta_y + \\
 + \epsilon \left [\frac{\omega_x'(\lambda)}{\omega_x(\lambda)} \right .&  \left . I_y\sin\theta_y\cos\theta_y+  \frac{\omega_y'(\lambda)}{\omega_y(\lambda)}I_x\sin\theta_x\cos\theta_x\right ] \, .
    \end{split}
    \label{eq:hamapp2}
\end{equation}
By applying the same transformation~\eqref{eq:genfun} one obtains
\begin{equation}
    \begin{split}
        H_{I_x \leftrightarrow I_y}(\bm \phi, \bm J,\lambda) & =-\delta(\lambda) \Ja+\omega_y \Jb+
\frac{\pi G}{{2 \omega_x(\lambda)\omega_y(\lambda)}} \times \\
& \times {\Ja (\Jb-\Ja)}\cos2 \phia+
O(\epsilon^2) \, .
    \end{split} 
    \label{eq:for_referee1}
\end{equation}
In this case, the amplitude detuning terms read $ \alpha_{xx}I_y^2 + 2 \alpha_{xy}I_x I_y + \alpha_{yy} I_x^2$, and following the same steps one arrives at the following model 
\begin{equation}
\begin{split}
    \ham_{I_x \leftrightarrow I_y} (\phia, & \Ja, \Jb) = -\delta (\lambda) \Ja + \frac{1}{2} \halfaa \Ja^2 + \alfab \Ja \Jb + \\
    + & G \Ja (\Jb-\Ja) \cos 2\phia + \qty[\omega_y(\lambda) \Jb + \frac{1}{2}\halfbb\Jb^2] \, ,
\end{split}
\label{eq:ham12app}
\end{equation}
where 
\begin{equation}
\begin{split}
   \frac{1}{2} \halfaa       & = \alpha_{xx}-2\alpha_{xy}+\alpha_{yy},\\
   \halfab             & =2\alpha_{xy}-2\alpha_{xx},\\
   \frac{1}{2} \halfbb & = \alpha_{xx} \, ,
\end{split}
\label{eq:ampdet2}
\end{equation}
where the second and third terms differ from the corresponding expressions in Eq.~\eqref{eq:ampdet1}. The previous Hamiltonian can be put into final form by dropping the terms in square brackets, \ie
\begin{equation}
\begin{split}
    \ham_{I_x \leftrightarrow I_y} (\phia, \Ja, \Jb) & = \left [ -\delta (\lambda) + \halfab \Jb \right ] \Ja + \frac{1}{2} \alfaa \Ja^2 + \\
    + & G \Ja (\Jb-\Ja) \cos 2\phia \\
    & = -\left \{ \left [ \delta (\lambda) - \halfab \Jb \right ] \Ja - \frac{1}{2} \alfaa \Ja^2 + \right . \\
    + & \left . G \Ja (\Jb-\Ja) \cos 2\phia \right \} \, ,
\end{split}
\label{eq:ham13app}
\end{equation}
where the initial phase of the angle has been shifted according to $\phia \to \phia + \pi/2$. This Hamiltonian should be compared with the original one, \ie without the exchange of the transverse planes, namely
\begin{equation}
\begin{split}
    \ham (\phia, & \Ja, \Jb) = \left [ \delta (\lambda) + \alfab \Jb \right ] \Ja + \frac{1}{2} \alfaa \Ja^2 + \\
    + & G \Ja (\Jb-\Ja) \cos 2\phia \, .
\end{split}
\label{eq:ham13}
\end{equation}

The two Hamiltonians differ by a global sign and a time-independent term linear in $\Ja$. This means that the equation of motions for the actions have the same form, and only the angles differ by a term linear in the time variable, which is irrelevant for the resonance crossing process, and hence $P_\text{na}$ remains invariant.

The other symmetry to be studied is that visible in Fig.~\ref{fig:axy_delta_k_n_pna} (bottom left). In this case, $P_\text{na}$ is quasiinvariant under the transformation
\begin{equation}
    \alpha\to -\alpha,\qquad K_3 \to -K_3 \, .
    \label{eq:symmetry1}
\end{equation}

Here, we recall the Hamiltonian system under consideration, namely
\begin{equation}
    \mathcal{H'}(\phi,J) = \eta J + \alpha J^2 + J (1-J)\cos 2\phi \, ,
    \label{eq:hamfapp}
\end{equation}
together with the two model parameters, \ie
\begin{equation} 
\eta=\frac{\delta+\alfab\Jb}{G\Jb},\qquad  \alpha=\frac{\alfaa}{2 \, G} \, ,
\label{eq:transformationapp}
\end{equation}
and we will express the Eqs.~\eqref{eq:detuning_map} and~\eqref{eq:alpha_map} in the following form
\begin{equation}
    \begin{split}
        \alfaa & = \gamma K_3 - \xi \, \overline{\alpha_{xy}} \, ,\\
        \alfab & = \hat{\gamma} K_3 + \hat{\xi} \, \overline{\alpha_{xy}}\, ,\\ 
        \alpha & = \bar{\gamma} -\bar{\xi} \, \frac{\overline{\alpha_{xy}}}{K_3}\,.
    \end{split}
     \label{eq:detuning_mapapp}
\end{equation}
knowing that $G \propto K_3$. Let us assume that the Hamiltonian~\eqref{eq:hamfapp} is considered for a set of parameters $\alpha^\ast$ and $K_3^\ast$. The following expressions can be easily derived
\begin{equation}
    \begin{split}
        \overline{\alpha_{xy}}\,^\ast & = \frac{\bar{\gamma}-\alpha^\ast}{\bar{\xi}} K_3^\ast \, , \\
        \eta^\ast & = \frac{\delta + \alfab^\ast \Jb}{\Jb G^\ast} \, , \\
        \alfab^\ast & = \left [ \hat{\gamma} - \frac{\hat{\xi} \left ( \alpha^\ast - \bar{\gamma} \right )}{\bar{\xi}} \right ] K_3^\ast \, .
    \end{split}
\end{equation}

Let us apply the symmetry~\eqref{eq:symmetry1}, \ie $\alpha^\ast \to -\alpha^\ast \, ,  \; K_3^\ast \to - K_3^\ast$. The value of the model parameters in this case is given by 
\begin{equation}
    \begin{split}
        \overline{\alpha_{xy}}\,^{\ast \ast} & = \frac{\bar{\gamma}-\alpha^\ast}{\bar{\xi}} \vert K_3^\ast \vert \, , \\
        \eta^{\ast \ast}   & = - \frac{\delta + \alfab^{\ast \ast} \Jb}{\Jb \vert G^\ast \vert } \, , \\
        \alfab^{\ast \ast} & = \left [ - \hat{\gamma} - \frac{\hat{\xi} \left ( \alpha^\ast - \bar{\gamma} \right )}{\bar{\xi}} \right ] \vert K_3^\ast \vert \, .
    \end{split}
\end{equation}
We observe that, while $\eta^{\ast \ast}$ seems to be the opposite of $\eta^\ast$, which is what is needed to satisfy the symmetry~\eqref{eq:symmetry0}, in fact, $\alfab^{\ast \ast} \neq \alfab^\ast$ from which we conclude that the symmetry is only partially satisfied, in particular, if $\vert \bar{\gamma} \vert \ll \vert \hat{\xi} (\alpha^\ast - \bar{\gamma}) / \bar{\xi}\vert $. 
\newpage
\bibliographystyle{unsrt}
\bibliography{mybibliography}
\end{document}